\documentclass{article}

\usepackage[utf8]{inputenc}
\usepackage[T1]{fontenc}

\PassOptionsToPackage{table, dvipsnames}{xcolor}
\usepackage{xcolor}

\PassOptionsToPackage{colorlinks=true, citecolor=blue, linkcolor=blue, urlcolor=black}{hyperref}

\usepackage{arxiv}
\usepackage[utf8]{inputenc}
\usepackage[T1]{fontenc}
\usepackage{url}
\usepackage{booktabs}
\usepackage{nicefrac}
\usepackage{microtype}
\usepackage{lipsum}
\usepackage{graphicx}
\usepackage{natbib}
\usepackage{doi}
\usepackage{algorithm}
\usepackage{algpseudocode}%
\usepackage{listings}%
\usepackage{multirow}
\usepackage{multicol}
\usepackage{amsmath,amssymb,amsfonts}
\usepackage{amsthm}
\usepackage{mathrsfs}
\usepackage{float}
\usepackage{bm}
\usepackage{caption}
\usepackage{hyperref}

\usepackage{tabularx}
\usepackage{array}
\usepackage{pdflscape}

\title{Incorporating wave physical priors into diffusion models: \\A novel approach to seismic resolution enhancement
}

\author{
	{Huanhuan~Tang, ~Weijian~Mao} \\
	Research Center for Computational and Exploration Geophysics, \\State Key Laboratory of Precision Geodesy,\\
    Innovation Academy for Precision Measurement Science and Technology, \\ Chinese Academy of Sciences, Wuhan 430077, China \\
        \And
	\href{https://orcid.org/0000-0001-8868-7967}{\includegraphics[scale=0.06]{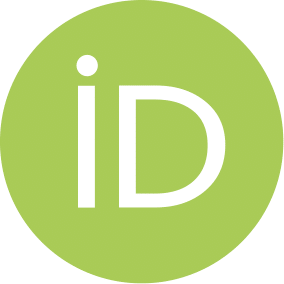}\hspace{1mm}Shijun~Cheng} \\
	Division of Physical Science and Engineering\\
	King Abdullah University of Science and Technology\\
	Thuwal 23955-6900, Saudi Arabia \\
        \And
    {Haoran~Zhang} \\
	School of Petroleum, China University of Petroleum-Beijing at Karamay, Xinjiang 834000, China \\
        \And
    {Yingying~Zhang} \\
	School of Geography and Tourism, Zhengzhou Normal University, Zhengzhou 450044, China \\
    [3ex]
  $^{*}$Corresponding author: \textbf{Shijun Cheng}~(\texttt{sjcheng.academic@gmail.com})
}

\renewcommand{\shorttitle}{PG-SSDM for seismic resolution enhancement}

\hypersetup{
pdftitle={A template for the arxiv style},
pdfsubject={q-bio.NC, q-bio.QM},
pdfauthor={David S.~Hippocampus, Elias D.~Striatum},
pdfkeywords={First keyword, Second keyword, More},
}

\begin{document}
\maketitle

\begin{abstract}
Seismic resolution enhancement remains a critical challenge in exploration geophysics, particularly when processing field data characterized by limited bandwidth, strong noise, and insufficient labeled training samples. Existing deep learning methods typically rely on supervised learning with synthetic training data, leading to distribution mismatch and poor generalization on real seismic acquisitions. To address these limitations, we develop a physics-guided self-supervised diffusion model (PG-SSDM) that learns directly from field observations without requiring paired high-resolution labels. The proposed framework combines three key innovations. First, a self-supervised training strategy constructs learning targets by progressively filtering the observed data itself, eliminating the need for high-resolution ground truth through iterative refinement across multiple stages. Second, seismic convolution model is embedded as a hard physical constraint in both the training loss function and the reverse sampling process, ensuring that generated high-resolution outputs respect fundamental seismic wave propagation physics. Third, the probabilistic nature of diffusion models enables uncertainty quantification, providing spatial confidence maps that identify regions where resolution enhancement may be less reliable. We validate PG-SSDM on synthetic data under various noise conditions and on a 3D post-stack field dataset. Experimental results demonstrate that the proposed method effectively recovers thin layers and subtle structures, suppresses noise, preserves structural continuity, thereby significantly improving the resolution and interpretability of seismic data.
\end{abstract}

\keywords{Seismic resolution enhancement \and Physics-guided \and Generative diffusion model \and Self-supervised learning.}
\section{Introduction}\label{sec:introduction}
Advancing exploration targets now commonly involve complex geological features, like thin interbeds, minor faults, and subtle reservoirs, which demand increasingly higher seismic data resolution. Unfortunately, field seismic data often fails to meet these interpretation requirements due to multiple factors: wave attenuation during propagation, equipment precision limitations, environmental noise, and earth filtering effects. While recent broadband seismic sources and high-density acquisition technologies represent significant progress, their high cost and operational constraints in the field mean that post-acquisition processing must compensate for hardware limitations \citep{2012Design}.

Existing approaches to enhance seismic resolution can be grouped into four main categories: deconvolution, inverse Q filtering, spectral whitening, and deep learning methods.

Deconvolution, a classical technique for enhancing seismic data resolution, aims to broaden the frequency band by compressing the seismic wavelet \citep{1988Least, 2006Principles, 1969PREDICTIVE}. Traditional methods, such as Wiener filtering \citep{2006Principles}, least-squares deconvolution \citep{1988Least}, predictive deconvolution \citep{1969PREDICTIVE}, and surface-consistent deconvolution \citep{ulvmoen2010improved}, rely on the convolution model and assume sparse reflection coefficients. However, these approaches assume locally stationary media and struggle when reflection coefficients are not sparse under complex geology. To address these limitations, \citep{2014Robust} introduced L1-norm regularization for sparse-constrained deconvolution, while \citep{ulvmoen2010improved} developed a Bayesian framework that incorporates geostatistical information through prior distributions. Yet all these methods assume a stationary wavelet during propagation, resulting in poor performance when processing non-stationary seismic signals. This motivated non-stationary deconvolution techniques, such as the generalized time-variant deconvolution theory \citep{koehler1985use} and Gabor transform-based deconvolution methods \citep{margrave2011gabor, jin2017research}, which allow the seismic wavelet to vary with time, thereby better handling the non-stationarity of seismic signals. Despite these advances, deconvolution faces several persistent challenges in field applications: (1) accurate wavelet estimation remains difficult, (2) the minimum-phase wavelet assumption often breaks down, (3) attenuated high-frequency components cannot be recovered, and (4) the inverse problem requires regularization parameters that are typically chosen empirically.

Inverse Q filtering improves resolution by compensating for amplitude loss and phase distortion caused by subsurface absorption \citep{sheriff1975factors, wang2002stable}. \cite{zhang2007inverse} developed a stable wavefield-based algorithm. \cite{hargreaves1991inverse} proposed a phase-only inverse Q filtering method grounded in the wavefield extrapolation principle. \cite{yao2003forward} introduced a depth-domain inverse Q filtering technique that simultaneously performs amplitude compensation and phase correction in the depth domain. Inverse Q filtering involves wavefield extrapolation methods where explicit approaches are unstable, leading to unstable amplitude compensation. Therefore, \cite{wang2006inverse} derived inverse Q filtering formulas with gain-limiting constraints and stabilization methods based on wavefield extrapolation. To address the numerical instability of inverse Q-filter amplitude compensation, \cite{zhang2015stable} proposed a self-adaptive, time-variant gain-limit approach. In pre-stack domains, inverse Q-filtering is often hindered by the difficulty in estimating layer-specific traveltimes. \cite{liu2020oriented} addressed this by transforming attenuation factors from the \(\tau\text{--}p\) domain to the \(t\text{--}x\) domain using inverse warped mapping, stabilized via Taylor-expansion-based division. Nevertheless, practical application of inverse Q filtering remains two fundamental challenges. Accurate Q-value estimation is difficult to achieve, and errors propagate directly into compensation artifacts. Moreover, field data are often noisy, and the amplitude compensation process amplifies noise along with signal, leading to signal-to-noise ratio (SNR) degradation.

Spectral whitening balances the energy distribution across different frequencies to enhance high-frequency components and, thus, improve vertical resolution. The basic method involves transforming seismic signals into the frequency domain via Fourier transform, followed by amplitude compensation within the effective frequency range \citep{allen1977short, gao2009enhancing}. However, Fourier transform inherently lacks the ability to analyze local frequency characteristics of signals. Alternatives based on wavelet transform \citep{grossmann1990reading, pinnegar2003s, xu2022compact} and Hilbert-Huang transform \citep{huang1998empirical, battista2007application, siyuan2023challenges} better handle nonlinear and non-stationary signals. Nonetheless, while spectral whitening enhances vertical resolution by boosting high-frequency components, it may distort the original amplitude-phase relationships, affecting the reliability of subsequent inversion and attribute extraction for quantitative interpretation. Furthermore, parameters such as wavelet basis selection and filtering thresholds require manual tuning, heavily relying on experience, which may lead to over- or under-whitening.

Deep learning (DL) avoids the restrictive assumptions and manual parameter selection inherent in traditional methods \citep{lecun2015deep, mousavi2022deep, yu2021deep}, offering new approaches for seismic resolution enhancement. Existing work can be broadly categorized into two methodological frameworks. The first treats resolution enhancement as a reflectivity inversion problem. \cite{kim2018geophysical} used 1D CNNs for trace-by-trace high-resolution inversion, but this approach is sensitive to noise and lateral variations, producing unstable results. To improve robustness, \cite{gao2022incorporating} introduced structural constraints based on $t-x$ predictive error filters (PEF), and \cite{chen2021optimization} combined traditional regularization with CNNs to learn regularization parameters. These methods assume a known, stationary wavelet. Yet wavelets are typically unknown and time-varying in practice. While \cite{gao2024high} achieved high-resolution mapping from acoustic impedance to band-limited non-stationary seismic data using TV regularization, reliable wavelet estimation from field data remains challenging. The second considers resolution enhancement as low-to-high frequency extension. \cite{li2021deep} trained a U-Net to map low- to high-frequency components using synthetic data generated by convolving reflectivity models with multi-frequency Ricker wavelets. Similarly, \cite{gao2022incorporating, gao2023deep} built training sets from well-log data and incorporated structural constraints to improve reconstruction accuracy.

However, most DL-based methods rely on supervised learning with labeled training data, which creates a fundamental generalization challenge: models trained on synthetic data often perform poorly on field data due to distribution mismatch, while obtaining sufficient labeled field data is impractical. Several strategies have been developed to address this limitation. \cite{jo2022machine} reduced the synthetic-to-field gap by extracting time-varying wavelets directly from field data at different depths to construct more realistic training sets. Domain adaptation methods aim to narrow the gap between synthetic and field data by applying domain-specific transformations, such as correlation-based alignment, to bring their input distributions closer and improve generalization \citep{alkhalifah2022mlreal, zhang2022improving, yang2025seismic}. More recently, self-supervised learning (SSL) has emerged as a promising alternative that bypasses the need for labeled data entirely \citep{cheng2024self}. In realm of seismic resolution enhancement, \cite{chai2023geophysics} and \cite{wang2023structurally} embedded a convolutional model in the loss function for geophysics-guided SSL deconvolution. \cite{cheng2025self} proposed an SSL framework for seismic resolution enhancement that reformulates the task as high-frequency reconstruction and achieves robust generalization through iterative pseudo-label refinement.

An additional limitation is that these methods are based on deterministic models, which produce a single deterministic solution after training. In contrast, generative diffusion models (GDMs) enable probabilistic modeling, capturing the uncertainty inherent in geophysical inverse problems. GDMs have been successfully applied to seismic modeling \citep{cheng2025seismic}, processing \citep{durall2023deep, liu2024generative, cheng2025gsfm, wang2025self}, imaging \citep{shi2024generative}, and inversion \citep{wang2023prior, zhang2025well}. Compared to generative adversarial networks (GANs), GDMs avoid adversarial training, achieve more stable optimization through gradual denoising, prevent mode collapse, and demonstrate better hyperparameter robustness \citep{dhariwal2021diffusion}. For resolution enhancement, \cite{zhang2024seisresodiff} applied a denoising diffusion probabilistic model (DDPM) conditioned on low-resolution seismic data to reconstruct high-resolution images. Also, \cite{yu2025unsupervised} used GDMs to seismic resolution enhancement by treating seismic deconvolution as an unsupervised generation problem, where they employed diffusion posterior sampling to inject observed data into the reverse process. However, their models are trained on synthetic data and then applied to field data. This means the learned distribution remains that of synthetic data, not field data, the same generalization gap that affects deterministic methods persists.

Table \ref{tab:method_comparison} systematically compares the major seismic resolution enhancement approaches across key dimensions including underlying principles, assumptions, data requirements, and limitations. This comparison highlights several important trends in the evolution of resolution enhancement techniques. Traditional methods (deconvolution, inverse Q filtering, spectral whitening) offer computational efficiency and solid theoretical foundations but are limited by restrictive assumptions about wavelet stationarity, accurate prior knowledge requirements, and poor noise robustness. Supervised deep learning methods demonstrate superior performance in controlled settings but face critical generalization challenges due to synthetic-to-field distribution mismatch. Recent advances in self-supervised learning and generative models attempt to address these limitations. However, existing approaches either lack physical constraints (pure SSL methods) or still rely on synthetic training data (supervised GDMs), leaving a gap for methods that can combine field-adaptive learning, physical consistency, and uncertainty quantification.

To address the aforementioned challenges, this paper proposes a physics-guided self-supervised diffusion model (PG-SSDM) that trains directly on field data without requiring labels. The method has three key features. First, a novel self-supervised learning strategy is adopted, which requires only the original low-resolution data itself for training, effectively linking the training objective with the input data. Second, it incorporates seismic wavelet information and convolutional physics into the reverse diffusion sampling process, enforcing physical consistency with Robinson's convolution model. Finally, it employs a composite loss function combining pixel loss with physical loss and total variation regularization, ensuring that predictions match data statistics, satisfy convolution physics, and maintain good signal-to-noise ratios and lateral continuity. We validate PG-SSDM through numerical experiments on both synthetic and field data. Synthetic tests evaluate the method's robustness under various noise types and intensity levels. The approach is then applied to a 3D land post-stack dataset, demonstrating its effectiveness on real field data.

\begin{table}[p]
\centering
\caption{Comparison of seismic resolution enhancement methods}
\label{tab:method_comparison}
\scriptsize

\rotatebox{90}{%
\begin{minipage}{\textheight}
\centering
\setlength{\tabcolsep}{4pt}
\renewcommand{\arraystretch}{1.25}

\begin{tabularx}{\textheight}{
p{3.0cm}
>{\centering\arraybackslash}X
>{\centering\arraybackslash}X
>{\centering\arraybackslash}X
>{\centering\arraybackslash}X
>{\centering\arraybackslash}X
>{\centering\arraybackslash}X
}
\toprule
\textbf{Method Category} 
& \textbf{Core Principle} 
& \textbf{Key Assumptions} 
& \textbf{Prior Knowledge Required} 
& \textbf{Training Data Needed} 
& \textbf{Main Advantages} 
& \textbf{Main Limitations} \\
\midrule

\multicolumn{7}{l}{\textit{Traditional Methods}} \\
\midrule
\addlinespace[0.3em]

Stationary Deconvolution 
& Wavelet compression via inverse filtering 
& Stationary wavelet, sparse reflectivity, minimum phase 
& Wavelet estimation 
& None 
& Well-established theory; computationally efficient 
& Cannot recover attenuated frequencies; fails for non-stationary signals \\

Non-stationary Deconvolution 
& Time-variant wavelet compression 
& Time-varying wavelet, local stationarity 
& Time-varying wavelet estimation 
& None 
& Handles signal non-stationarity 
& Complex wavelet estimation; high computational cost \\

Inverse Q Filtering 
& Amplitude and phase compensation for absorption 
& Known Q model; stable extrapolation 
& Q-value distribution 
& None 
& Compensates physical attenuation 
& Q estimation difficult; noise amplification \\

Spectral Whitening 
& Frequency-domain energy balancing 
& Effective frequency band definition 
& Frequency range selection 
& None 
& Simple implementation 
& Distorts amplitude--phase relations; manual tuning required \\

\addlinespace[0.6em]
\midrule
\multicolumn{7}{l}{\textit{Deterministic Deep Learning Methods}} \\
\midrule
\addlinespace[0.3em]

Supervised Reflectivity Inversion 
& CNN-based inverse mapping 
& Known stationary wavelet 
& Wavelet; paired labels 
& High--low resolution pairs (synthetic) 
& Learns complex mappings 
& Poor field generalization; wavelet dependence \\

Supervised Frequency Extension 
& Low-to-high frequency mapping (U-Net) 
& Training and field distributions match 
& None (implicitly learned) 
& Multi-resolution synthetic data 
& Avoids explicit wavelet estimation 
& Synthetic-to-field gap; data-hungry \\

Domain Adaptation 
& Distribution alignment across domains 
& Transferable latent features exist 
& None 
& Synthetic data + limited field data 
& Reduces domain mismatch 
& Limited adaptation capacity \\

Self-Supervised Learning 
& Pseudo-label generation from data 
& Method-dependent 
& Method-dependent 
& None (field data only) 
& No labeled data; field adaptive 
& Weak physical constraints; iterative refinement \\

\addlinespace[0.6em]
\midrule
\multicolumn{7}{l}{\textit{Probabilistic Generative Models}} \\
\midrule
\addlinespace[0.3em]

Supervised GDMs 
& Conditional generative modeling 
& Distribution learned from training data 
& None 
& Paired synthetic data 
& Uncertainty quantification; stable training 
& Generalization limited by synthetic data \\

\textbf{Physics-Guided SSL GDMs (Proposed)} 
& Self-supervised probabilistic generation with physics constraints 
& Physical convolutional forward model 
& Wavelet estimate 
& None (self-supervised on field data) 
& Label-free; physically consistent; uncertainty-aware; field adaptive 
& Requires wavelet estimation \\

\bottomrule
\end{tabularx}
\end{minipage}%
}
\end{table}

\section{Method}\label{sec:method}

\subsection{Review of conditional diffusion model-based seismic resolution enhancement}\label{md:review}
\begin{figure}[!t] 
\centering
\includegraphics[width=\textwidth]{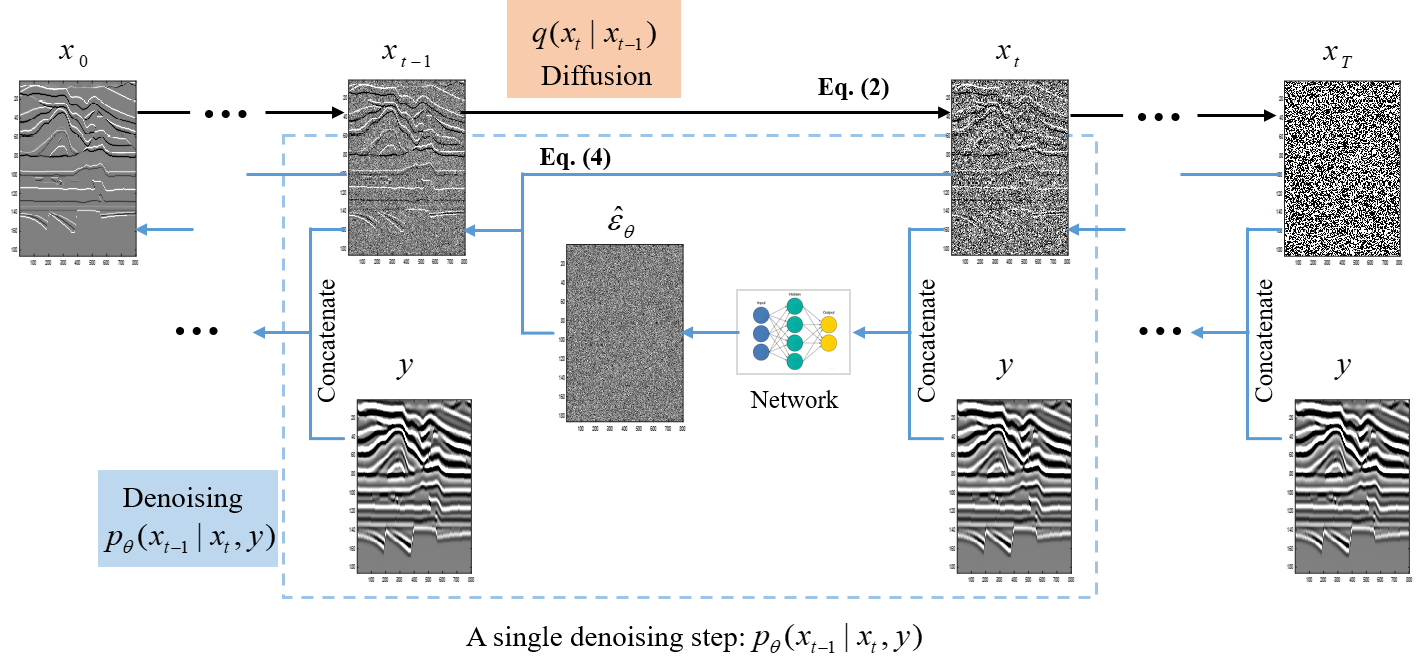}
\caption{Diagram of the conditional diffusion model-based seismic resolution enhancement \citep{zhang2024seisresodiff}.}
\label{fig1}
\end{figure}

GDMs \citep{ho2020denoising} have recently been adopted for seismic resolution enhancement due to their ability to learn powerful generative priors that map low-resolution seismic data to high-resolution representations. In this setting, a diffusion model aims to learn the conditional distribution $p(x_0 \mid \tilde{x})$, where $x_0$ denotes the desired high-resolution seismic profile and $\tilde{x}$ is its degraded or band-limited counterpart. The conditional diffusion approaches is illustrated in Figure \ref{fig1}. In this framework, the forward diffusion process gradually corrupts $x_0$ with Gaussian noise following a fixed variance schedule, producing noisy variables $x_t$ via
\begin{equation}\label{eq1}
q(x_t \mid x_{t-1})=\mathcal{N}\!\left(x_t;\sqrt{1-\beta_t}\,x_{t-1},\,\beta_t\mathbf{I}\right),
\end{equation}
and equivalently,
\begin{equation}\label{eq2}
x_t=\sqrt{\bar{\alpha}_t}\,x_0+\sqrt{1-\bar{\alpha}_t}\,\epsilon ,
\end{equation}
where $\epsilon\sim\mathcal{N}(0,\mathbf{I})$ represents Gaussian noise and $\bar{\alpha}_t=\prod_{i=1}^t (1-\beta_i)$. This corruption process is deterministic and requires no learning.

Resolution enhancement is achieved during the reverse process, where the model reconstructs \(x_0\) from noisy inputs while being guided by the low-resolution constraint \(\tilde{x}\). A neural network (NN) is trained to approximate the reverse transition distribution
\begin{equation}\label{eq3}
p_\theta(x_{t-1}\mid x_t,\tilde{x})=
\mathcal{N}\!\left(x_{t-1};\mu_\theta(x_t,t,\tilde{x}),\,\Sigma_\theta(x_t,t)\right),
\end{equation}
where conditioning on \(\tilde{x}\) enforces structural consistency between the generated reflectivity and the observed band-limited data. In practice, most seismic super-resolution approaches adopt the noise-prediction parameterization, where the network receives \((x_t,\tilde{x})\) as input and predicts the added noise \(\epsilon_\theta(x_t,t,\tilde{x})\). The reverse update then takes the form
\begin{equation}\label{eq4}
x_{t-1}
=\frac{1}{\sqrt{\alpha_t}}
\left(
x_t-\frac{1-\alpha_t}{\sqrt{1-\bar{\alpha}_t}}\,
\epsilon_\theta(x_t,t,\tilde{x})
\right)
+\sigma_t z,
\end{equation}
which iteratively removes noise until a high-resolution estimate \(x_0\) is obtained, where $z\sim\mathcal{N}(0,\mathbf{I})$. This conditional diffusion mechanism leverages the robustness of the denoising process and ensures that large-scale structures represented in \(\tilde{x}\) are preserved, while the generative prior synthesizes the missing high-frequency components. However, existing conditional diffusion approaches rely on paired high-low-resolution seismic data for supervised training and require repeated noise-prediction across diffusion steps, resulting in high computational cost and limiting their applicability in field scenarios where high-resolution labels are unavailable. 

These limitations motivate the development of alternative training strategies that do not depend on paired supervision and that more efficiently parameterize the reverse process. In the following subsection, we introduce a self-supervised diffusion framework that constructs its own training pairs from observed data and adopts an \(x_0\)-prediction formulation, enabling survey-specific adaptation without requiring clean high-resolution ground truth.


\subsection{Self-supervised diffusion models for seismic resolution enhancement}\label{md:ssldiff}

Building on the limitations of conditional diffusion models discussed above, we develop a diffusion framework that leverages the observed seismic data itself to drive the learning of high-resolution reconstructions. Instead of relying on externally provided paired high–low-resolution examples, the method derives both the conditioning input and the supervision target directly from the acquired seismic data. This is achieved by generating a physically meaningful low-resolution counterpart through a low-pass operator and constructing noisy diffusion samples via the forward corruption process. The network then learns to predict the clean seismic data \(x_0\) from these self-generated pairs, enabling the model to adapt naturally to the survey-specific spectral content and structural characteristics without requiring any high-resolution ground truth.

The framework begins by generating a physically meaningful low-resolution version of the observed seismic data \(x\) through a low-pass operator \(\mathcal{F}_{\mathrm{LP}}\), producing
\begin{equation}\label{eq5}
\tilde{x}=\mathcal{F}_{\mathrm{LP}}(x).
\end{equation}
This low-pass filtering operation attenuates high-frequency components in the seismic data, effectively reducing its resolution. By applying such filtering to the original field data \(x\), we deliberately create a lower-resolution version \(\tilde{x}\) that serves as a controllable degradation. This strategy enables us to construct self-supervised training pairs where the original data \(x\) acts as the target distribution and the filtered data \(\tilde{x}\) provides the conditioning constraint. In this manner, the diffusion model can be trained to learn the conditional distribution \(p(x|\tilde{x})\) without requiring external paired labels, a key advantage inspired by the self-supervised denoising framework of \cite{cheng2025self}.

Meanwhile, applying the forward diffusion process to \(x\) produces noisy samples
\begin{equation}\label{eq6}
x_t=\sqrt{\bar{\alpha}_t}\,x+\sqrt{1-\bar{\alpha}_t}\,\epsilon, \qquad \epsilon\sim\mathcal{N}(0,\mathbf{I}),
\end{equation}
forming self-generated training triplets \(\{(x_t,\tilde{x}),x\}\). Unlike standard diffusion models that predict the noise \(\epsilon\), our network directly predicts the clean high-resolution data as follows
\begin{equation}\label{eq7}
\hat{x}_0 = f_\theta(x_t, \tilde{x}, t),
\end{equation}
where \(f_\theta\) is the U-Net-based conditional generator.

In the initial stage, the network is trained using the self-generated pairs \(\{(x_t,\tilde{x}),x\}\) with a multi-objective loss function:
\begin{equation}\label{eq8}
\mathcal{L}_{\text{init}}(\theta)=
\mathbb{E}_{t,\,\epsilon}\!
\left[
\mathcal{L}_{\text{MSE}}(f_\theta, x_t, \tilde{x}, x, t) + \mathcal{L}_{\text{reg}}(f_\theta, x_t, \tilde{x}, t)
\right],
\end{equation}
where \(\mathcal{L}_{\text{MSE}}\) represents the reconstruction loss and \(\mathcal{L}_{\text{reg}}\) encompasses physics-guided regularization terms. The detailed formulation of each loss component will be presented in the next subsection. The network is trained for \(N_{\text{iter}}\) iterations, acquiring a basic capability to enhance seismic resolution while respecting physical constraints.

After the initial training stage, the framework proceeds through multiple progressive stages to approximate increasingly higher-resolution conditional distributions. Specifically, at the end of the initial stage, the trained network generates an enhanced output:
\begin{equation}\label{eq9}
y^{(1)}=f_{\theta_{\text{init}}}(x, \tilde{x}, t{=}1),
\end{equation}
where \(\theta_{\text{init}}\) denotes the parameters after initial training. To construct the first refinement stage, we generate a new low-resolution version
\begin{equation}\label{eq10}
\tilde{x}^{(1)} = \mathcal{F}_{\mathrm{LP}}(y^{(1)}),
\end{equation}
and apply forward diffusion to \(y^{(1)}\):
\begin{equation}\label{eq11}
y_t^{(1)}=\sqrt{\bar{\alpha}_t}\,y^{(1)}+\sqrt{1-\bar{\alpha}_t}\,\epsilon.
\end{equation}
This forms new training triplets \(\{(y_t^{(1)}, \tilde{x}^{(1)}), y^{(1)}\}\) with a stronger resolution contrast between the conditioning input \(\tilde{x}^{(1)}\) and the target output \(y^{(1)}\).

The network is then retrained using the same multi-objective loss function structure:
\begin{equation}\label{eq12}
\mathcal{L}^{(1)}(\theta)=
\mathbb{E}_{t,\,\epsilon}\!
\left[
\mathcal{L}_{\text{MSE}}(f_\theta, y_t^{(1)}, \tilde{x}^{(1)}, y^{(1)}, t) + \mathcal{L}_{\text{reg}}(f_\theta, y_t^{(1)}, \tilde{x}^{(1)}, t)
\right],
\end{equation}
again for \(N_{\text{iter}}\) iterations. This process repeats for \(K\) progressive stages. At stage \(k\), the enhanced output from the previous stage serves as the new training target:
\begin{equation}\label{eq13}
y^{(k)}=f_{\theta^{(k-1)}}(y^{(k-1)}, \tilde{x}^{(k-1)}, t{=}1),
\end{equation}
with the corresponding low-resolution condition
\begin{equation}\label{eq14}
\tilde{x}^{(k)} = \mathcal{F}_{\mathrm{LP}}(y^{(k)}),
\end{equation}
and diffused samples
\begin{equation}\label{eq15}
y_t^{(k)}=\sqrt{\bar{\alpha}_t}\,y^{(k)}+\sqrt{1-\bar{\alpha}_t}\,\epsilon.
\end{equation}

Through these progressive stages, the model iteratively learns to map increasingly degraded low-resolution inputs to progressively enhanced high-resolution targets, gradually approximating the conditional distribution of higher-resolution seismic data. The complete training procedure is summarized in Algorithm~\ref{alg1}.

\begin{algorithm}[t]
\caption{Physics-guided self-supervised diffusion training}
\label{alg1}
\begin{algorithmic}[1]
\Require Observed low-resolution seismic data $x$, low-pass operator $\mathcal{F}_{\mathrm{LP}}$, diffusion schedule $\{\beta_t\}_{t=1}^T$, number of progressive stages $K$, iterations per stage $N_{\text{iter}}$
\Ensure Trained enhancement model $f_\theta$
\State
\State \textbf{// Initial training stage}
\State $\tilde{x}^{(0)} \leftarrow \mathcal{F}_{\mathrm{LP}}(x)$ \Comment{Generate initial low-resolution constraint}
\State $y^{(0)} \leftarrow x$ \Comment{Initial target is the observed data}
\For{$i = 1$ to $N_{\text{iter}}$}
    \State Sample $t \sim \mathrm{Uniform}\{1,\ldots,T\}$
    \State Sample $\epsilon \sim \mathcal{N}(0,\mathbf{I})$
    \State $x_t = \sqrt{\bar{\alpha}_t}\,y^{(0)} + \sqrt{1-\bar{\alpha}_t}\,\epsilon$ 
    \State $\hat{x}_0 = f_\theta(x_t,\tilde{x}^{(0)},t)$ \Comment{Predict high-resolution output}
    \State Compute $\mathcal{L}_{\text{init}}(\theta)$ with $\mathcal{L}_{\text{MSE}}$ and $\mathcal{L}_{\text{reg}}$ \Comment{Multi-objective loss}
    \State Update $\theta$ by gradient descent on $\mathcal{L}_{\text{init}}(\theta)$
\EndFor
\State $\theta_{\text{init}} \leftarrow \theta$ \Comment{Save initial stage parameters}
\State
\State \textbf{// Progressive refinement stages}
\For{$k = 1$ to $K$}
    \State $y^{(k)} \leftarrow f_{\theta}(y^{(k-1)}, \tilde{x}^{(k-1)}, t{=}1)$ \Comment{Generate enhanced output from previous stage}
    \State $\tilde{x}^{(k)} \leftarrow \mathcal{F}_{\mathrm{LP}}(y^{(k)})$ \Comment{Create new low-resolution condition}
    \For{$i = 1$ to $N_{\text{iter}}$}
        \State Sample $t \sim \mathrm{Uniform}\{1,\ldots,T\}$
        \State Sample $\epsilon \sim \mathcal{N}(0,\mathbf{I})$
        \State $y_t^{(k)} = \sqrt{\bar{\alpha}_t}\,y^{(k)} + \sqrt{1-\bar{\alpha}_t}\,\epsilon$
        \State $\hat{y}_0^{(k)} = f_\theta(y_t^{(k)},\tilde{x}^{(k)},t)$
        \State Compute $\mathcal{L}^{(k)}(\theta)$ with $\mathcal{L}_{\text{MSE}}$ and $\mathcal{L}_{\text{reg}}$ \Comment{Multi-objective loss}
        \State Update $\theta$ by gradient descent on $\mathcal{L}^{(k)}(\theta)$
    \EndFor
\EndFor
\State
\State \Return trained enhancement model $f_\theta$
\end{algorithmic}
\end{algorithm}

\subsection{Physics-guided composite loss function}\label{md:pgloss}

As mentioned in the previous subsection, both the initial training stage and the progressive refinement stages employ a multi-objective loss function to guide network optimization. This composite loss integrates data fidelity, physical constraints, sparsity priors, and spatial regularity to ensure that the enhanced seismic data maintains high resolution while respecting physical authenticity and geological plausibility.

The total loss function at stage $k$ (including the initial stage with $k=0$) is formulated as:
\begin{equation}\label{eq16}
\mathcal{L}^{(k)}(\theta) = \lambda_{\text{mse}} \mathcal{L}_{\text{mse}}^{(k)} + \lambda_{\text{phy}} \mathcal{L}_{\text{phy}}^{(k)} + \lambda_{L_1} \mathcal{L}_{L_1}^{(k)} + \lambda_{\text{tv}} \mathcal{L}_{\text{tv}}^{(k)},
\end{equation}  
where $\mathcal{L}_{\text{mse}}^{(k)}$, $\mathcal{L}_{\text{phy}}^{(k)}$, $\mathcal{L}_{L_1}^{(k)}$, and $\mathcal{L}_{\text{tv}}^{(k)}$ represent the mean squared error loss, physical constraint loss, $L_1$ sparsity loss, and total variation (TV) regularization, respectively. The hyperparameters $\lambda_{\text{mse}}$, $\lambda_{\text{phy}}$, $\lambda_{L_1}$, and $\lambda_{\text{tv}}$ control the relative importance of each term and remain constant across all training stages.

The mean squared error (MSE) loss serves as the foundational reconstruction objective, minimizing the distance between the predicted high-resolution output and the target data:
\begin{equation}\label{eq17}
\mathcal{L}_{\text{mse}}^{(k)} = \mathbb{E}_{t,\epsilon} \left[ \| y^{(k)} - f_\theta(y_t^{(k)}, \tilde{x}^{(k)}, t) \|_2^2 \right],
\end{equation}  
where $y^{(k)}$ denotes the target data at stage $k$ (with $y^{(0)} = x$ for the initial stage), $y_t^{(k)}$ is the corresponding noisy sample from forward diffusion, $\tilde{x}^{(k)} = \mathcal{F}_{\mathrm{LP}}(y^{(k)})$ is the low-resolution conditioning input, and $f_\theta$ is the neural network. This term ensures basic reconstruction fidelity.

The physics-guided loss represents the key component of our method, incorporating the seismic convolution model as a physical constraint:
\begin{equation}\label{eq18}
\mathcal{L}_{\text{phy}}^{(k)} = \mathbb{E}_{t,\epsilon} \left[ \| \tilde{x}^{(k)} -f_\theta(y_t^{(k)}, \tilde{x}^{(k)}, t) \ast w \|_2^2 \right],
\end{equation}
where $w$ is the known or estimated seismic wavelet and $\ast$ denotes convolution. This loss enforces that the predicted high-resolution output, after convolution with the wavelet, should match the low-resolution input $\tilde{x}^{(k)}$, ensuring consistency with Robinson's convolution model. By back-propagating gradients of $\mathcal{L}_{\text{phy}}$ through the network parameters $\theta$, the generation process is guided toward physically plausible solutions that conform to seismic wave propagation physics.

The $L_1$ sparsity loss promotes sparse reflection coefficients, consistent with geological priors:
\begin{equation}\label{eq19}
\mathcal{L}_{L_1}^{(k)} = \mathbb{E}_{t,\epsilon} \left[ \| f_\theta(y_t^{(k)}, \tilde{x}^{(k)}, t) \|_1 \right].
\end{equation}
Sparsity encourages sharper seismic events and clearer boundaries, enhancing the visual resolution of geological structures.

The TV loss promotes spatial smoothness by penalizing differences between horizontally adjacent samples:
\begin{equation}\label{eq20}
\mathcal{L}_{\text{tv}}^{(k)} = \mathbb{E}_{t,\eta} \left[ \frac{1}{H \times (W - 1)} \sum_{i=1}^{H} \sum_{j=1}^{W-1} \sqrt{(\hat{x}_{0,i,j+1}^{(k)} - \hat{x}_{0,i,j}^{(k)})^2 + \eta} \right],
\end{equation}
where $\hat{x}_0^{(k)} = f_\theta(y_t^{(k)}, \tilde{x}^{(k)}, t)$, $H$ and $W$ are the height and width of the output, and $\eta$ is a small constant for numerical stability. This term improves lateral continuity of seismic signals and mitigates the negative impact of random or anomalous noise on resolution enhancement.

By integrating these four complementary objectives, the composite loss function ensures that the diffusion model learns to enhance seismic resolution while maintaining physical consistency with the convolution model, geological plausibility through sparsity, spatial coherence through TV regularization, and overall fidelity through MSE reconstruction. This multi-faceted approach is essential for producing high-quality, geologically interpretable enhanced seismic data.

\subsection{Physics-guided reverse sampling process}\label{md:pgsampling}

While the physics-guided composite loss function enforces physical consistency during training, we further integrate physical constraints into the reverse sampling (inference) process to ensure that the generated high-resolution seismic data respects Robinson's convolution model at each denoising step. This is achieved through the physics-guided posterior sampling (PGPS) strategy, where we leverages the physical loss $\mathcal{L}_{\text{phy}}$ as a guidance to boost the sampling process.

Recall that in the self-supervised framework described in Section \ref{md:ssldiff}, the network directly predicts the clean high-resolution data $\hat{x}_0$ from the noisy sample $x_t$ at each timestep $t$. Given this predicted $\hat{x}_0$, the standard reverse diffusion process computes the posterior mean based on Bayes' rule:
\begin{equation}\label{eq21}
\mu_\theta(x_t, \tilde{x}, t) = \frac{\sqrt{\bar{\alpha}_{t-1}} \beta_t}{1 - \bar{\alpha}_t} \hat{x}_0 + \frac{\sqrt{\alpha_t}(1 - \bar{\alpha}_{t-1})}{1 - \bar{\alpha}_t} x_t,
\end{equation}
where $\hat{x}_0 = f_\theta(x_t, \tilde{x}, t)$ and $\alpha_t = 1 - \beta_t$. The reverse diffusion step then proceeds as:
\begin{equation}\label{eq22}
x_{t-1} = \mu_\theta(x_t, \tilde{x}, t) + \sigma_t z,
\end{equation}
where $\sigma_t$ is the noise scale, and $z \sim \mathcal{N}(0, \mathbf{I})$ for $t > 1$, $z = 0$ for $t = 1$.

To guide the reverse sampling toward physically plausible solutions, we incorporate Robinson's convolution model as a hard constraint. Robinson's model describes the relationship between observed seismic data, wavelet, and reflectivity:
\begin{equation}\label{eq23}
x = x_0 \ast w,
\end{equation}
where $x_0$ represents the high-resolution reflectivity (the desired output), $w$ is the seismic wavelet, and $x$ is the original observed low-resolution seismic data. 

At each reverse diffusion timestep $t$, the network predicts $\hat{x}_0^{(t)} = f_\theta(x_t, \tilde{x}, t)$ from the current noisy sample $x_t$. To ensure this prediction satisfies the physical constraint, we compute the gradient of the physical loss with respect to $x_t$:
\begin{equation}\label{eq24}
\nabla_{x_t} \mathcal{L}_{\text{phy}}(x_t) = \nabla_{x_t} \| x - \hat{x}_0^{(t)} \ast w \|_2^2.
\end{equation}
This gradient quantifies how the current prediction $\hat{x}_0^{(t)}$ deviates from the physical constraint in Eq.~(\ref{eq23}). Therefore, we can use this gradient to correct the intermediate sample $x_t$ before applying the standard reverse diffusion step:
\begin{equation}\label{eq25}
\tilde{x}_{t} = x_{t} - \lambda \cdot \nabla_{x_t} \mathcal{L}_{\text{phy}}(x_t),
\end{equation}
where $\lambda$ is the guidance strength parameter controlling the magnitude of the physical correction, and $\tilde{x}_t$ denotes the physics-corrected intermediate sample. This correction steers $x_t$ toward regions of the latent space where the convolution model is better satisfied.

The corrected sample $\tilde{x}_t$ is then used in the standard reverse diffusion update. Specifically, we re-predict the clean data from the corrected sample $\hat{x}_0^{\text{corr}} = f_\theta(\tilde{x}_t, \tilde{x}, t)$ and compute the posterior mean based on the direct $x_0$ prediction parameterization. The reverse diffusion step becomes:
\begin{equation}\label{eq26}
x_{t-1} = \frac{\sqrt{\bar{\alpha}_{t-1}} \beta_t}{1 - \bar{\alpha}_t} f_\theta(\tilde{x}_t, \tilde{x}, t) + \frac{\sqrt{\alpha_t}(1 - \bar{\alpha}_{t-1})}{1 - \bar{\alpha}_t} \tilde{x}_t + \sigma_t z.
\end{equation}
The first two terms in Eq.~(\ref{eq26}) represent the posterior mean $\mu_\theta(\tilde{x}_t, \tilde{x}, t)$, which combines the network's prediction of the clean data with the current corrected noisy sample to estimate the previous timestep's sample mean.

This two-step correction mechanism ensures that at each denoising iteration, the sampling trajectory is first adjusted to align with physical constraints, and then the standard diffusion dynamics drive the sample toward higher-resolution data. The negative gradient $-\lambda \cdot \nabla_{x_t} \mathcal{L}_{\text{phy}}(x_t)$ acts as a correction force derived from the physics of seismic wave propagation. Conceptually, the network first generates a candidate high-resolution prediction based on learned priors, and the physical guidance then refines the intermediate state to ensure compatibility with the known wavelet and low-resolution input. 

The physics-guided reverse sampling process is summarized in Algorithm~\ref{alg2}. By integrating physical guidance into both training (via $\mathcal{L}_{\text{phy}}$ in the composite loss) and inference (via PGPS gradient correction), our framework ensures that the enhanced seismic data maintains physical authenticity throughout the entire generation pipeline. As a result, we can enable the model to reconstruct high-resolution details that are both sharp and consistent with seismic physics.

\begin{algorithm}[t]
\caption{Physics-guided reverse sampling for inference}
\label{alg2}
\begin{algorithmic}[1]
\Require Observed low-resolution seismic data $x$, trained model $f_\theta$, wavelet $w$, diffusion schedule $\{\beta_t\}_{t=1}^T$, guidance strength $\lambda$
\Ensure High-resolution seismic data $x_0$
\State $x_T \sim \mathcal{N}(0, \mathbf{I})$ \Comment{Initialize with pure noise}
\For{$t = T, T-1, \ldots, 1$}
    \State $\hat{x}_0^{(t)} \leftarrow f_\theta(x_t, x, t)$ \Comment{Predict clean high-resolution data}
    \State Compute $\nabla_{x_t} \mathcal{L}_{\text{phy}}(x_t) = \nabla_{x_t} \| x - \hat{x}_0^{(t)} \ast w \|_2^2$
    \State $\tilde{x}_t \leftarrow x_t - \lambda \cdot \nabla_{x_t} \mathcal{L}_{\text{phy}}(x_t)$ \Comment{Physics-guided correction}
    \State $\hat{x}_0^{\text{corr}} \leftarrow f_\theta(\tilde{x}_t, x, t)$ \Comment{Re-predict from corrected sample}
    \State $\mu_\theta \leftarrow \frac{\sqrt{\bar{\alpha}_{t-1}} \beta_t}{1 - \bar{\alpha}_t} \hat{x}_0^{\text{corr}} + \frac{\sqrt{\alpha_t}(1 - \bar{\alpha}_{t-1})}{1 - \bar{\alpha}_t} \tilde{x}_t$ \Comment{Compute posterior mean}
    \If{$t > 1$}
        \State Sample $z \sim \mathcal{N}(0, \mathbf{I})$
        \State $x_{t-1} \leftarrow \mu_\theta + \sigma_t z$ \Comment{Reverse diffusion with noise}
    \Else
        \State $x_0 \leftarrow \mu_\theta$ \Comment{Final step without noise}
    \EndIf
\EndFor
\State \Return $x_0$
\end{algorithmic}
\end{algorithm}

\subsection{Uncertainty quantification}\label{md:uq}

A key advantage of the diffusion model framework, compared to deterministic DL approaches, is its inherent capability for probabilistic modeling. Unlike deterministic networks that produce a single fixed output for a given input, diffusion models generate samples from a learned conditional distribution $p(x_0|\tilde{x})$, naturally enabling uncertainty quantification, which represents a critical feature for assessing the reliability of enhanced results in geophysical applications.

During the reverse sampling process described before, the stochastic injection of Gaussian noise $z \sim \mathcal{N}(0, \mathbf{I})$ at each timestep $t$ introduces variability in the generation trajectory. By performing $M$ independent sampling runs with different noise realizations for the same low-resolution input $\tilde{x}$, we can obtain an ensemble of high-resolution outputs $\{x_0^{(m)}\}_{m=1}^M$, each representing a plausible solution consistent with both the conditioning input and the learned physical constraints.

The ensemble mean provides a stable estimate of the high-resolution data:
\begin{equation}\label{eq27}
\bar{x}_0 = \frac{1}{M} \sum_{m=1}^{M} x_0^{(m)},
\end{equation}
where $x_0^{(m)}$ denotes the $m$-th generated sample. More importantly, the ensemble standard deviation quantifies spatial uncertainty:
\begin{equation}\label{eq28}
\sigma_{x_0}(i,j) = \sqrt{\frac{1}{M} \sum_{m=1}^{M} (x_0^{(m)}(i,j) - \bar{x}_0(i,j))^2},
\end{equation}
where $(i,j)$ indexes spatial positions. High values of $\sigma_{x_0}$ indicate regions where the model exhibits greater variability across samples, suggesting lower confidence in the enhanced high-resolution content.

This uncertainty map serves multiple practical purposes. First, it identifies geologically complex or poorly constrained regions, like such as thin-layer boundaries, complex fault zones, and discontinuous reservoirs, where resolution enhancement becomes more challenging. Second, it provides interpreters with a quantitative measure of confidence for assessing the reliability of recovered high-frequency components. Third, it enables risk-aware decision making in subsequent interpretation workflows, such as reservoir characterization and drilling site selection, by flagging areas that may require additional data acquisition or alternative interpretation strategies.

\subsection{Network architecture}\label{md:net}

The neural network $f_\theta$ is based on the U-Net architecture from improved denoising diffusion probabilistic models \citep{nichol2021improved}, adapted for seismic resolution enhancement. We employ a 5-level U-Net with an encoder-decoder structure. The encoder extracts multi-scale features through progressive downsampling, while the decoder reconstructs high-resolution seismic details through upsampling. The network configuration is as follows: the initial channel count is 64, with channel multipliers of $(1, 2, 4, 8, 8)$ at each level, yielding 64, 128, 256, 512, and 512 channels at resolutions $1, 1/2, 1/4, 1/8,$ and $1/16$ of the input size, respectively. Skip connections between corresponding encoder and decoder layers preserve fine-scale details during reconstruction. To capture long-range dependencies in geological structures, we incorporate self-attention mechanisms at the bottleneck layer and at resolutions $1/8$ and $1/16$, which enables the network to model spatial relationships across distant locations. The network uses sinusoidal positional embeddings for timestep $t$ and concatenates the low-resolution condition $\tilde{x}$ with the noisy input $x_t$ along the channel dimension.
\section{Numerical Examples}\label{examples}

\subsection{Synthetic data with Gaussian noise}

\begin{figure}[!t] 
\centering
\includegraphics[width=\textwidth]{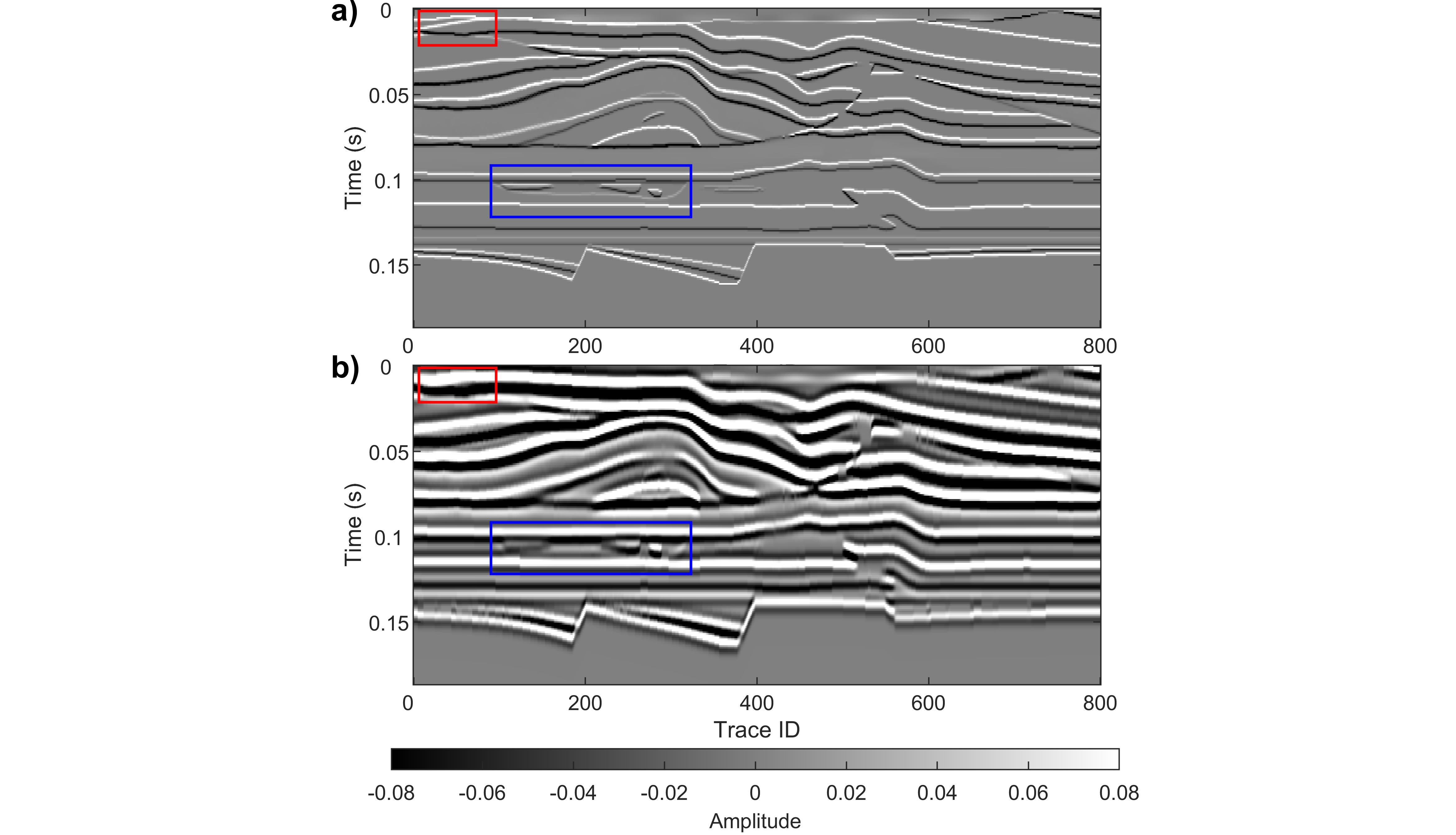}
\caption{Reflectivity and synthetic post-stack seismic data of the Overthrust model.}
\label{fig2}
\end{figure}

We first evaluate the proposed method on synthetic data contaminated with Gaussian noise to assess its fundamental performance and noise robustness. The vertical reflectivity model is derived from the Overthrust velocity-density model, as shown in Figure \ref{fig2}(a). We convolve this reflectivity with a Ricker wavelet of 80 Hz dominant frequency to generate low-resolution seismic data (Figure \ref{fig2}(b)). The resulting synthetic dataset comprises 801 traces, each containing 186 time samples at a 1 ms sampling interval. The band-limited nature of the seismic wavelet significantly reduces resolution, obscuring several weak reflectors present in the original reflectivity model. For example, the first reflector within the red rectangular region in the upper-left portion of Figure \ref{fig2}(a) becomes virtually indistinguishable in Figure \ref{fig2}(b) due to wavelet side-lobe interference. Similarly, subtle geological features within the blue rectangular area around 0.1 seconds exhibit blurred boundaries in Figure \ref{fig2}(b), making precise interpretation challenging.

To validate the robustness of the proposed method under realistic acquisition conditions, we contaminate the synthetic low-resolution seismic data with varying levels of Gaussian random noise. The noise is added following $d' = d + \lambda \cdot \text{std}(d) \cdot \mathcal{N}(0,1)$, where $d$ is the clean low-resolution data, $\text{std}(d)$ is its standard deviation, $\mathcal{N}(0,1)$ denotes standard Gaussian noise, and $\lambda$ controls the noise intensity. We create four datasets with $\lambda = 0.0, 0.1, 0.3,$ and $0.5$, representing progressively increasing noise levels, which serve as original observed data for resolution enhancement test.

\begin{figure*}[!b] 
\centering
\includegraphics[width=\textwidth]{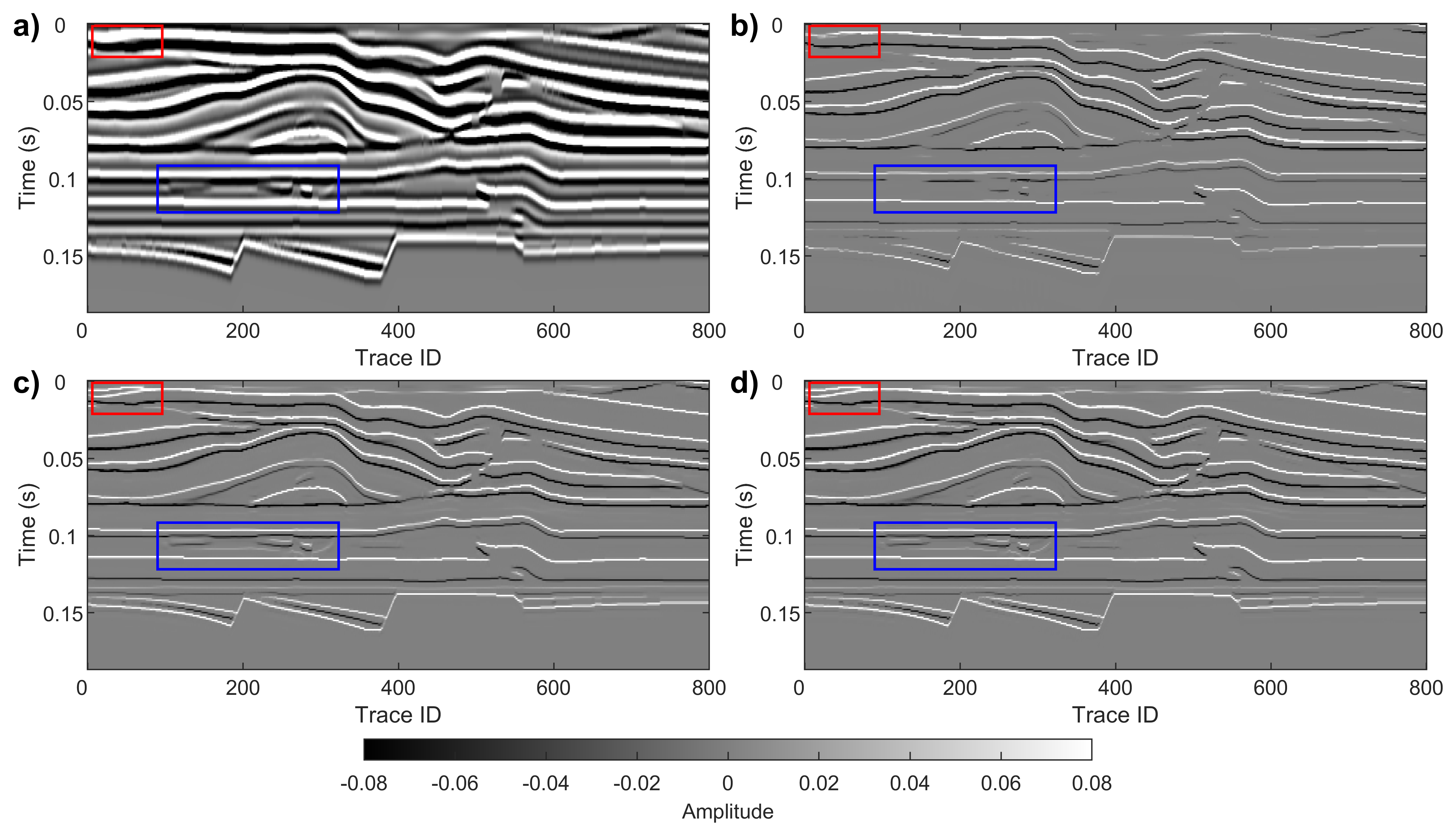}
\caption{Noise-free synthetic data test. (a) Original low-resolution data. (b) SSI method result. (c) PISSD prediction. (d) Proposed method prediction.}
\label{fig3}
\end{figure*}

\begin{figure*}[!t] 
\centering
\includegraphics[width=\textwidth]{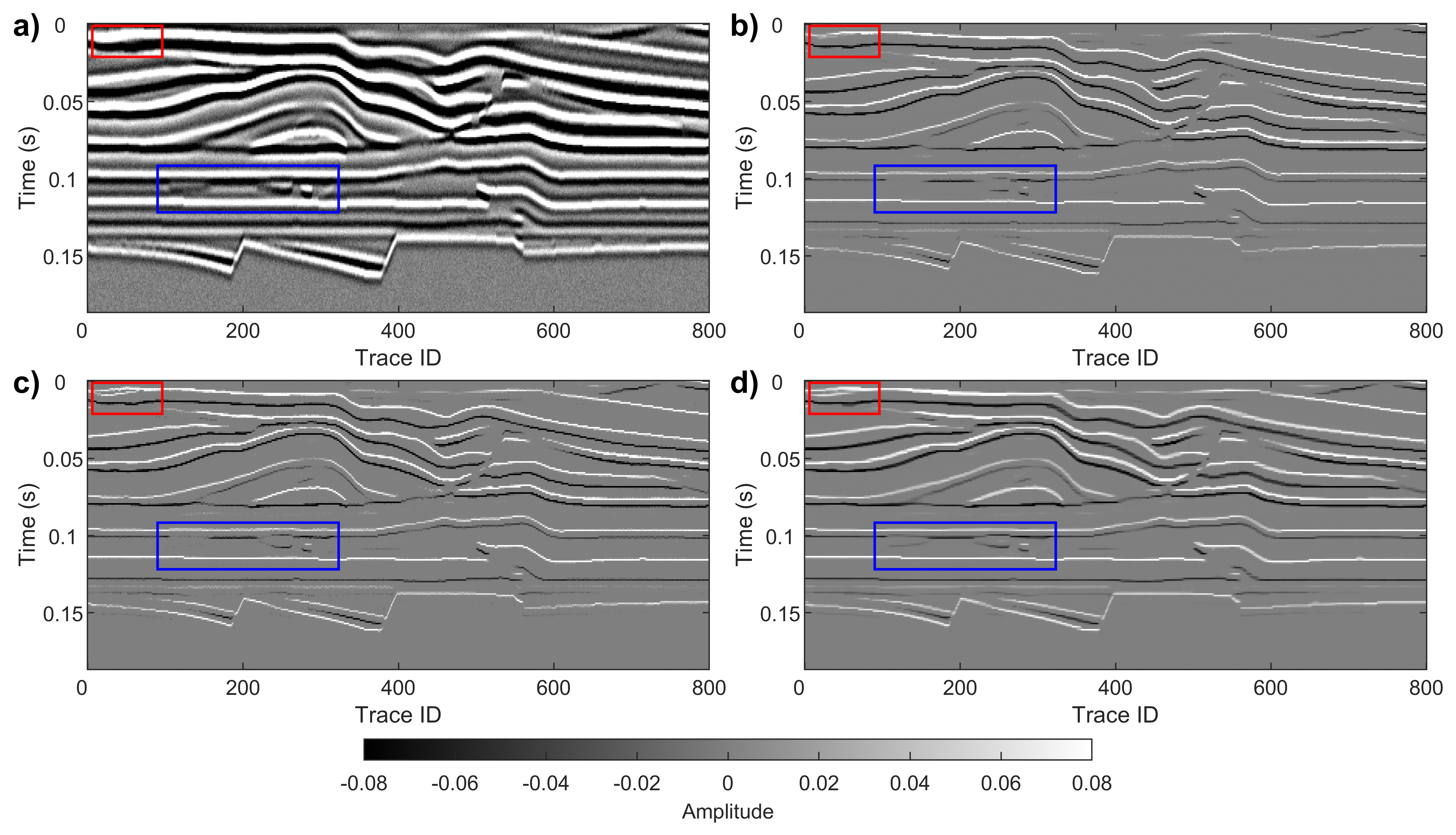}
\caption{Noisy synthetic data test (Noise level $ = 0.1$). (a) Original low-resolution data. (b) SSI method result. (c) PISSD prediction. (d) Proposed method prediction.}
\label{fig4}

\vspace{0.2cm}
 
\centering
\includegraphics[width=\textwidth]{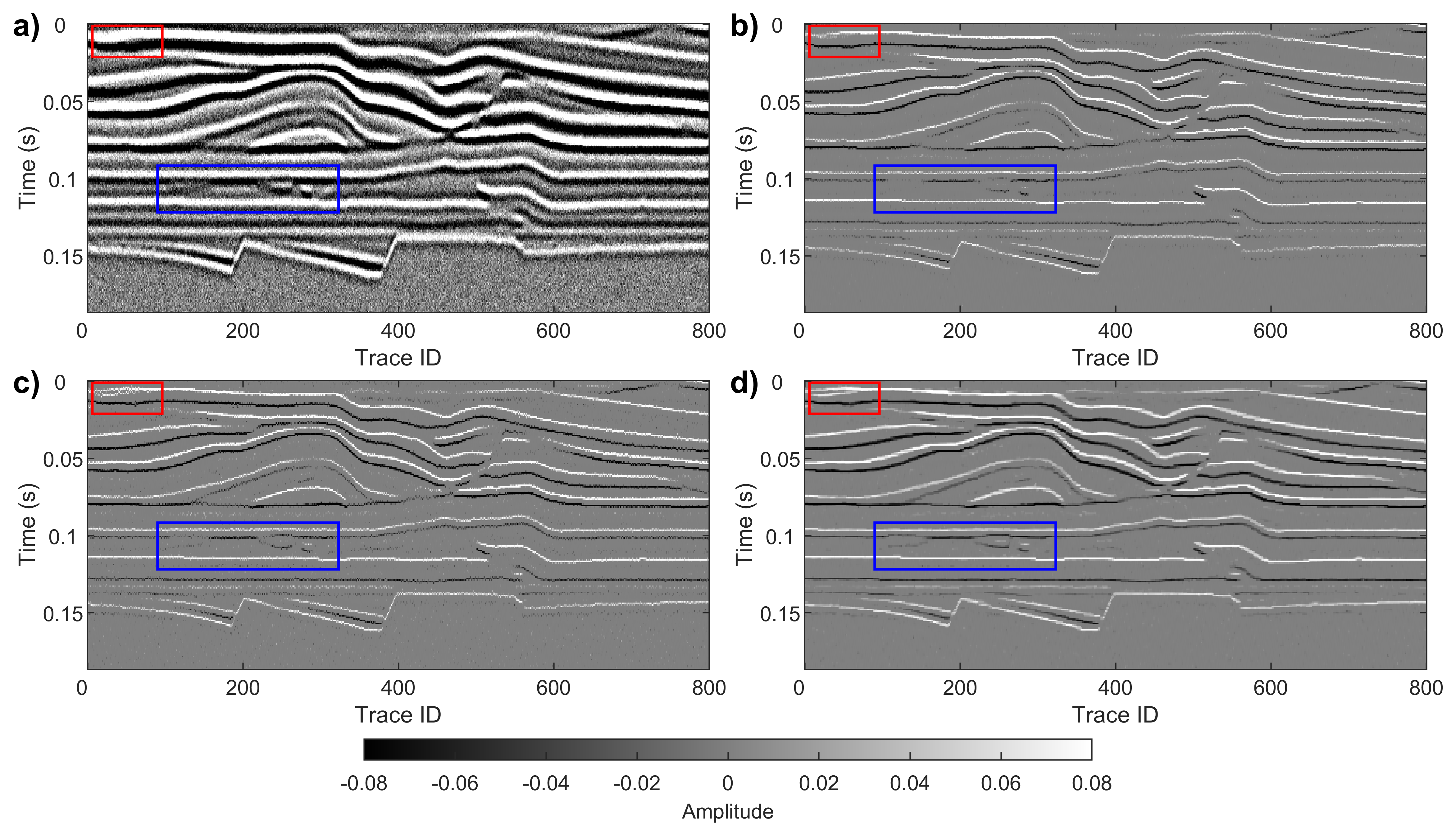}
\caption{Noisy synthetic data test (Noise level $ = 0.3$). (a) Original low-resolution data. (b) SSI method result. (c) PISSD prediction. (d) Proposed method prediction.}
\label{fig5}
\end{figure*}

\begin{figure*}[!t] 
\centering
\includegraphics[width=\textwidth]{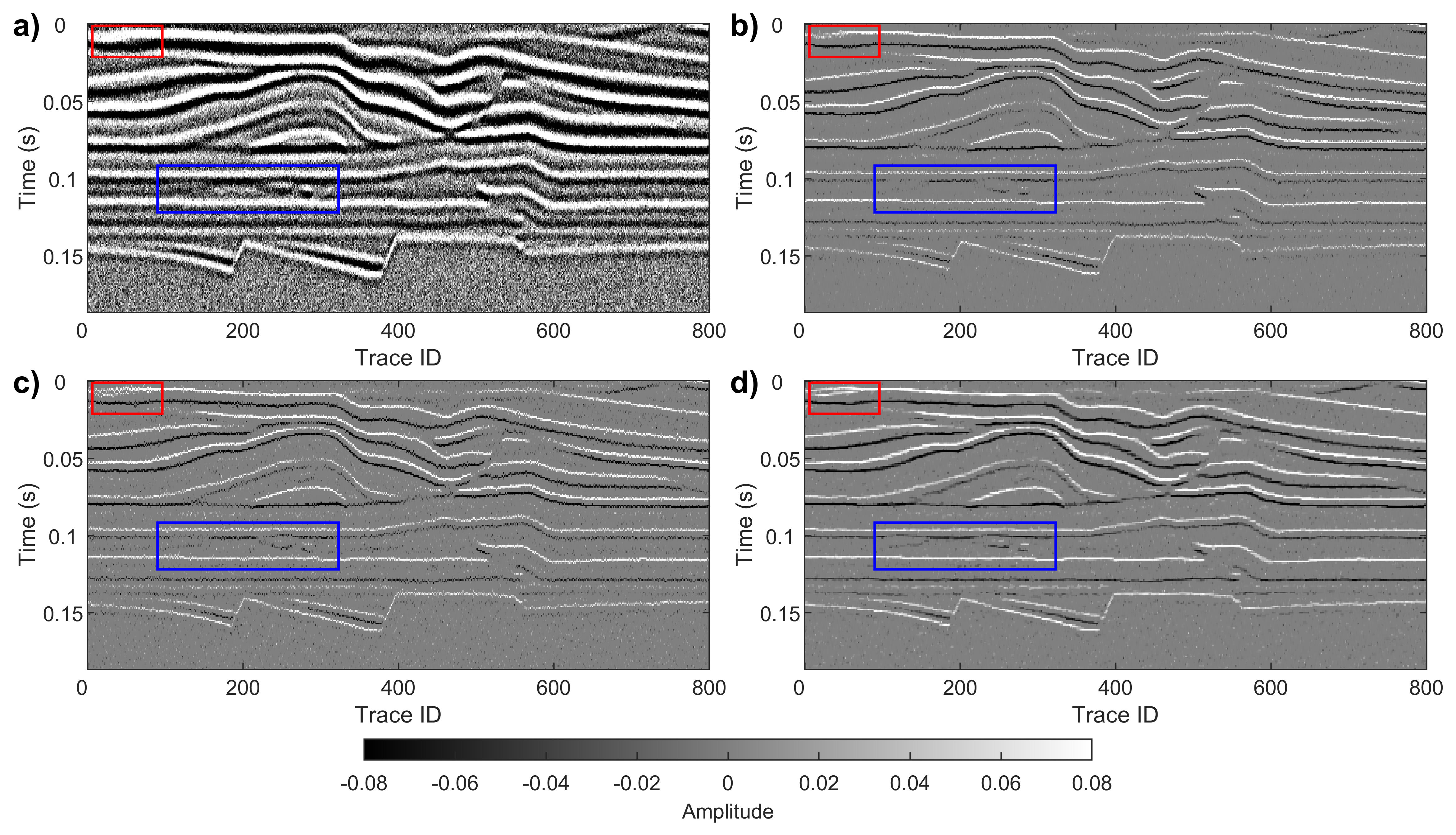}
\caption{Noisy synthetic data test (Noise level $ = 0.5$). (a) Original low-resolution data. (b) SSI method result. (c) PISSD prediction. (d) Proposed method prediction.}
\label{fig6}
\end{figure*}

We compare the proposed method against two benchmark approaches: sparse spike inversion (SSI) \citep{2014Robust} and physics-informed self-supervised deconvolution (PISSD) \citep{chai2023geophysics}. To ensure fair comparison and isolate the contribution of our framework from architectural differences, both the proposed method and PISSD employ the same U-Net architecture described in Section \ref{md:net}. The PISSD implementation differs only in that it removes the timestep embedding and uses a single-channel input consisting solely of the observed seismic data, rather than the multi-channel noisy-clean pair used in our framework. Figures \ref{fig3} to \ref{fig6} compare the resolution enhancement performance of the three methods across the four noise levels ($\lambda = 0.0, 0.1, 0.3,$ and $0.5$). In each figure, subplot (a) shows the input low-resolution data, while subplots (b), (c), and (d) present results from SSI, PISSD, and the proposed method, respectively.

Under noise-free conditions (Figure \ref{fig3}), all three methods improve data resolution to varying degrees. However, in the region highlighted by the red rectangle, the SSI result still exhibits noticeable wavelet side-lobe effects, producing composite events that merge adjacent weak reflections without clear separation. Both PISSD and the proposed method achieve more effective side-lobe suppression. For the subtle geologic features indicated by the blue rectangle, the SSI method fails to provide discernible structural recovery, whereas PISSD and the proposed method successfully resolve such fine details.

As noise level increases ($\lambda \geq 0.1$, Figures \ref{fig4} to \ref{fig6}), the performance differences become more pronounced. The SSI method consistently struggles to distinguish closely spaced weak reflectors, as seen in the red rectangular zones. In the blue rectangular region near 0.1 seconds, where the original low-resolution data show ambiguous structural boundaries, SSI offers limited improvement and introduces high-frequency artifacts due to its inherent noise sensitivity. PISSD exhibits some denoising capability but tends to oversmooth subtle stratigraphic features. In contrast, the proposed method maintains superior performance across all noise levels, recovering sharp structural boundaries, suppressing noise effectively, and preserving stratigraphic continuity.

Table \ref{tab1} provides quantitative comparisons using root mean square error (RMSE) and structural similarity index measure (SSIM) metrics across different noise levels. The proposed method achieves the lowest RMSE and highest SSIM under all noise conditions, demonstrating superior overall performance. Under noise-free conditions, the proposed method performs comparably to PISSD, with both significantly outperforming SSI. This highlights the inherent advantages of deep learning approaches in resolution enhancement tasks. As noise level increases, this performance gap becomes significant. At $\lambda = 0.5$, the RMSE values of SSI and PISSD rise to 0.0184 and 0.0194, respectively, while the proposed method increases only marginally to 0.0145. For SSIM, both SSI and PISSD exhibit notable degradation, whereas the proposed method maintains a high value of 0.8932. These results validate the exceptional noise robustness and stability of the proposed method, confirming its suitability for real-world seismic data characterized by random noise contamination.

\begin{table*}[!t]
\centering
\caption{Performance comparison under different noise levels}
\label{tab1}
\begin{tabular}{lcccccc}
\toprule
\multirow{2}{*}{Noise Level} & \multicolumn{3}{c}{RMSE} & \multicolumn{3}{c}{SSIM} \\
\cmidrule(lr){2-4} \cmidrule(lr){5-7}  
 & SSI & PISSD & \textbf{Our Method} & SSI & PISSD & \textbf{Our Method} \\
\midrule
0.0 & 0.0157 & 0.0104 & \textbf{0.0101} & 0.8819 & 0.9553 & \textbf{0.9599} \\
0.1 & 0.0158 & 0.0145 & \textbf{0.0132} & 0.8804 & 0.9044 & \textbf{0.9168} \\
0.3 & 0.0168 & 0.0162 & \textbf{0.0134} & 0.8680 & 0.8815 & \textbf{0.9131} \\
0.5 & 0.0184 & 0.0194 & \textbf{0.0145} & 0.8434 & 0.8296 & \textbf{0.8932} \\
\bottomrule
\end{tabular}
\end{table*}

\begin{figure}[!b] 
\centering
\includegraphics[width=0.55\textwidth]{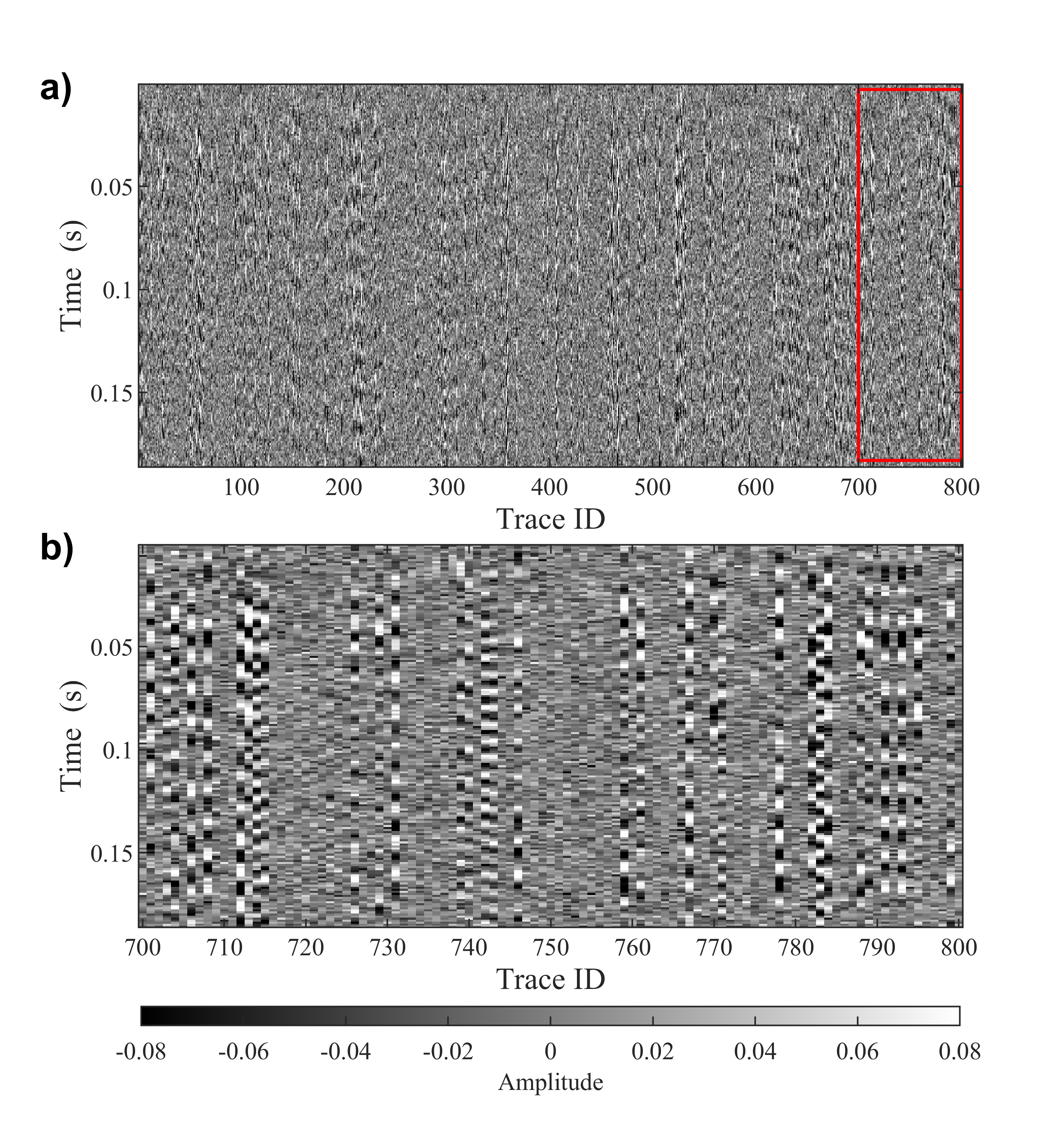}
\caption{The realistic noise extracted from field data. (a) Random noise and coupling noise. (b) Zoomed-in view corresponding to the red box area in (a).}
\label{fig7}
\end{figure}

\begin{figure*}[!t] 
\centering
\includegraphics[width=\textwidth]{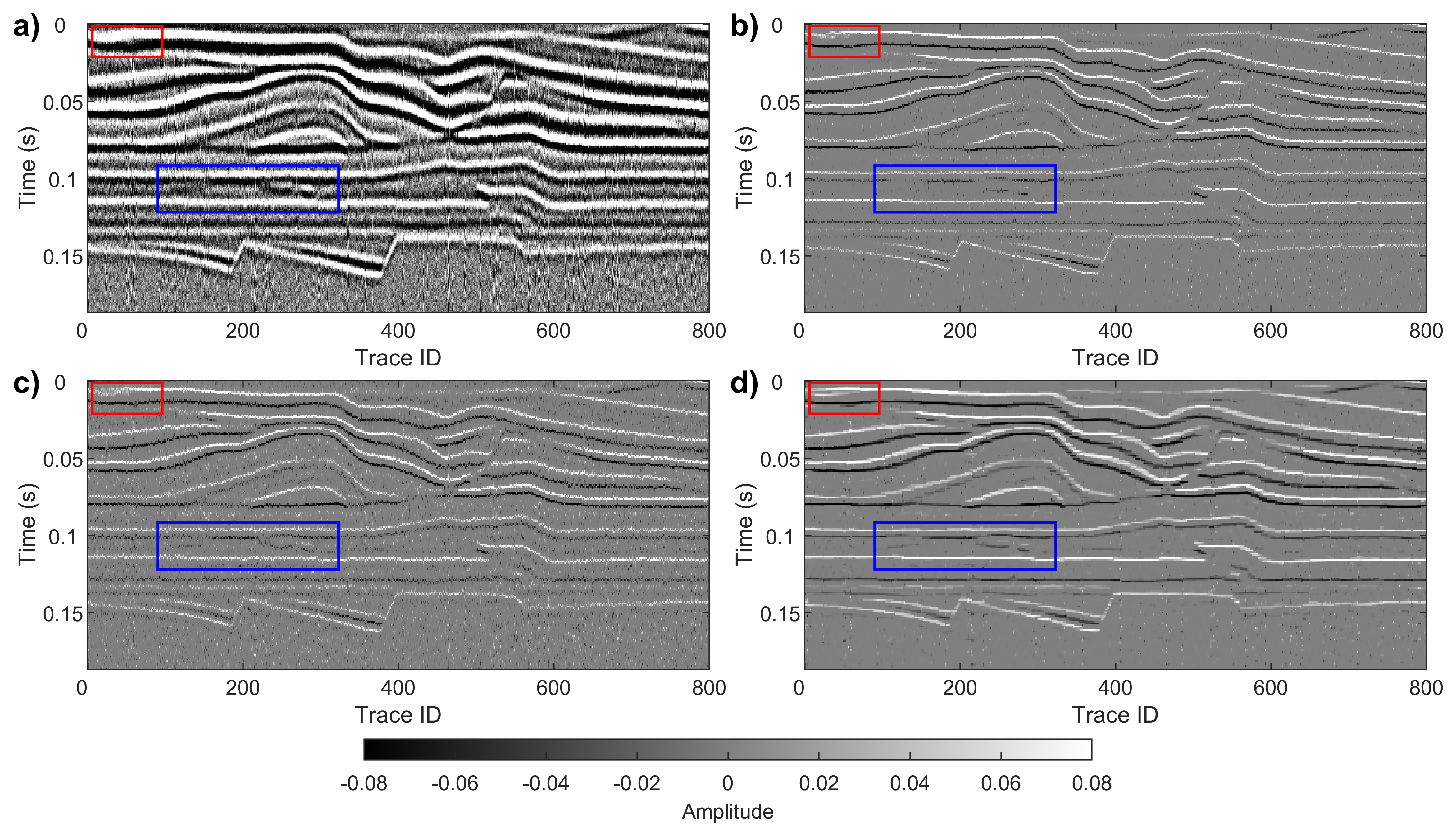}
\caption{Multi-noisy synthetic data test. (a) Original low-resolution data. (b) SSI method result. (c) PISSD prediction. (d) Proposed method prediction.}
\label{fig8}
\end{figure*}

\begin{figure*}[!b] 
\centering
\includegraphics[width=0.8\textwidth]{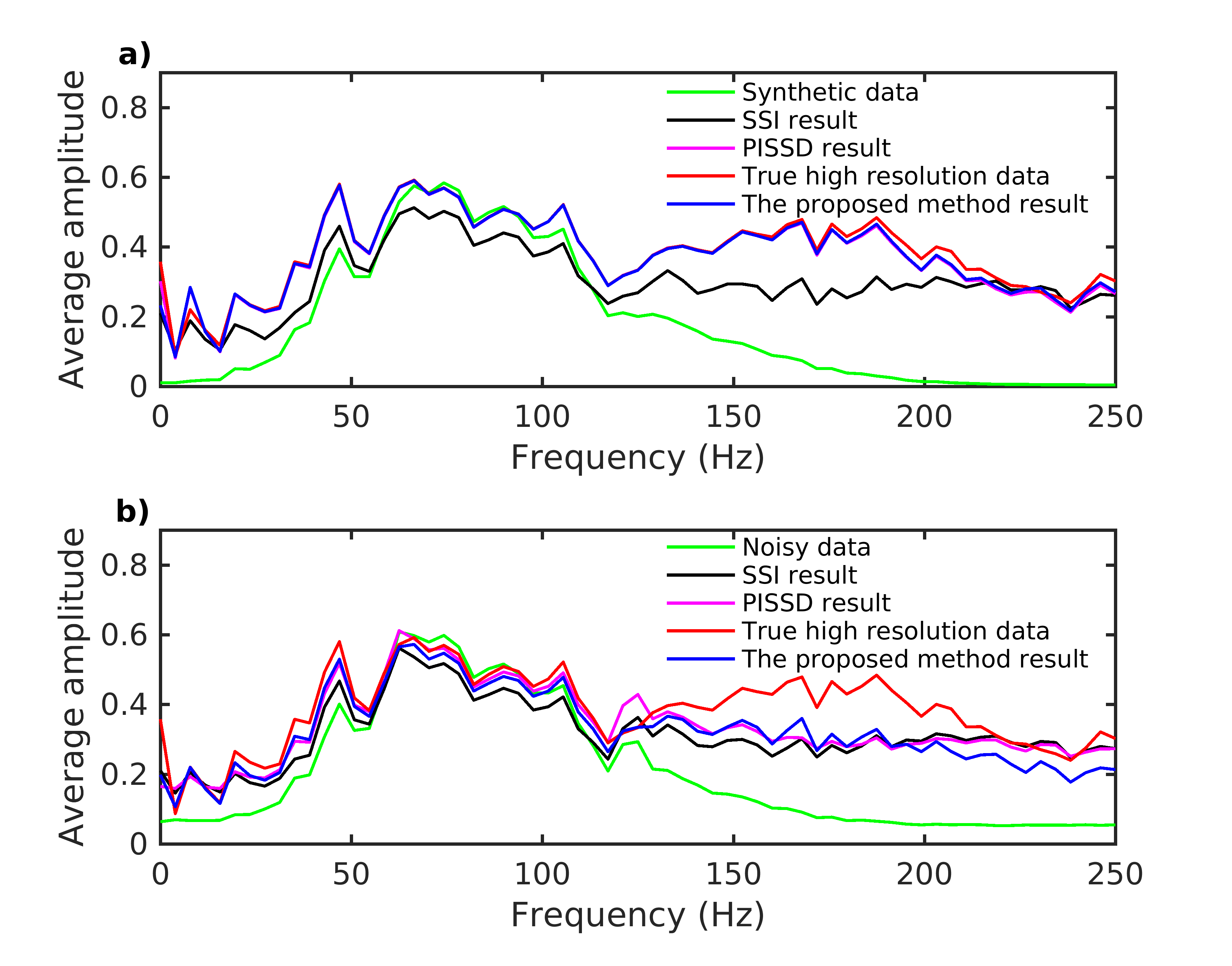}
\caption{Amplitude spectrum comparison between the original synthetic data, the ture high-resolution data, and the predictions from the three methods. (a) Noise-free condition. (b) Multi-noise condition.}
\label{fig9}
\end{figure*}

\subsection{Synthetic data with realistic noise}

While the previous section demonstrated the method's robustness under Gaussian random noise, field seismic data typically contain more complex interference patterns. Beyond random noise, various types of coherent noise are commonly present, such as the coupling noise prevalent in Distributed Acoustic Sensing (DAS) data. This type of noise significantly overlaps with the effective signal in the frequency domain, making it difficult to suppress without damaging the signal \citep{tang2025edcc}. To evaluate the proposed method under more realistic conditions, we introduce coupling noise extracted from real DAS data into the synthetic dataset, in addition to random Gaussian noise.

Figure \ref{fig7}(a) shows the coupling noise pattern extracted from field DAS data. Figure \ref{fig7}(b) presents a zoomed-in view of a local section, where the characteristic striping pattern of coupling noise is clearly distinguishable from random noise in both morphology and spatial distribution. We add this coupling noise to the synthetic low-resolution data along with Gaussian random noise ($\lambda = 0.3$) to simulate realistic field acquisition conditions. The contaminated seismic data is shown in Figure \ref{fig8}(a), with results from SSI, PISSD, and the proposed method displayed in Figure \ref{fig8}(b), (c), and (d), respectively.

We can see that in the red rectangular area, the proposed method effectively identifies and separates adjacent formation interfaces despite strong noise interference. In contrast, both SSI and PISSD fail to achieve clear stratigraphic differentiation, reflecting their limitations in simultaneous noise suppression and resolution enhancement. The blue rectangular region contains fine geological structures where signal continuity and structural integrity are critical. Here, the proposed method also outperforms the two benchmark approaches, further validating its capability to recover details in complex noise environments. Quantitative evaluation confirms these observations. The proposed method achieves RMSE of 0.0150 and SSIM of 0.8843, substantially outperforming SSI (RMSE: 0.0192, SSIM: 0.8248) and PISSD (RMSE: 0.0208, SSIM: 0.7786). These results demonstrate the method's advantages in both numerical reconstruction accuracy and structural fidelity under realistic noise conditions.

\begin{figure*}[!b] 
\centering
\includegraphics[width=\textwidth]{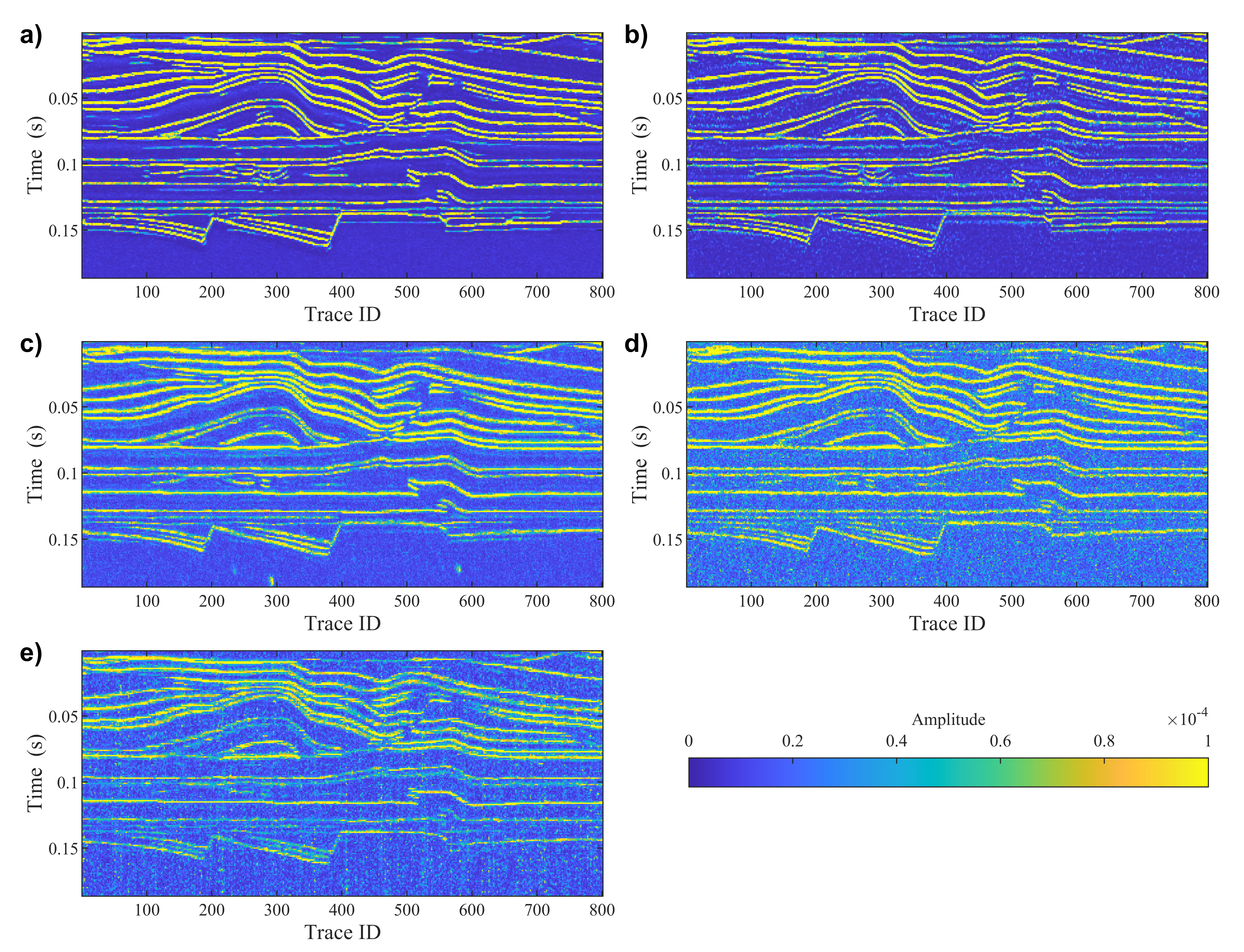}
\caption{Standard deviation map of the 10 predictions using our method. (a) $\lambda = 0.0$ profile. (b) $\lambda = 0.1$ profile. (c) $\lambda = 0.3$ profile. (d) $\lambda = 0.5$ profile. (e) Realistic noise profile.}
\label{fig10}
\end{figure*}

To further assess frequency enhancement capabilities, we compare the amplitude spectra of results from the three methods under both noise-free (Figure \ref{fig3}) and realistic noise conditions (Figure \ref{fig8}). Figures \ref{fig9}(a) and (b) show these comparisons. All methods substantially enhance high-frequency amplitude above 115 Hz, confirming their capability to extend the frequency bandwidth effectively. Under noise-free conditions, the amplitude spectra from the proposed method and PISSD closely align with the true high-resolution data, clearly surpassing SSI. In noisy environments, all methods exhibit some performance degradation. However, the proposed method maintains the smallest deviation from the true amplitude spectrum, highlighting its superior robustness in high-frequency reconstruction and noise resistance.


To demonstrate the uncertainty quantification capability of the proposed method introduced in Section \ref{md:uq},  we generate ten high-resolution realizations for each test case using different noise initializations while maintaining the same conditioning inputs (low-resolution seismic data and wavelet). Figure \ref{fig10} shows the standard deviation maps computed across these realizations for different noise scenarios: (a) noise-free, (b)-(d) Gaussian noise with levels $\lambda = 0.1, 0.3, 0.5$, and (e) realistic noise conditions. The uncertainty maps reveal several important characteristics:
\begin{itemize}
    \item First, noise significantly elevates prediction uncertainty. Compared to the noise-free case (Figure \ref{fig10}(a)), all noisy scenarios exhibit substantially higher uncertainty levels. As Gaussian noise intensity increases from $\lambda = 0.1$ to $0.5$ (Figures \ref{fig10}(b)-(d)), the overall uncertainty progressively intensifies, as indicated by the warmer colors spreading across the sections.
    \item Second, noise contamination fundamentally alters the spatial distribution of uncertainty. In the noise-free case (Figure \ref{fig10}(a)), uncertainty is primarily localized near reflection events, appearing as discrete patterns following the geological structures. However, as noise is introduced and its intensity increases, uncertainty becomes pervasive across the entire grid. This contamination effect is particularly evident in Figures \ref{fig10}(c)-(d), where elevated uncertainty values are distributed uniformly throughout the sections rather than being confined to specific geological features. The realistic noise case (Figure \ref{fig10}(e)) exhibits similar widespread uncertainty, with additional striping patterns reflecting the coherent coupling noise characteristics.
    \item Third, strong reflection events consistently correspond to higher uncertainty values. This pattern is observable across all scenarios, where zones with high-amplitude reflectors exhibit elevated standard deviation. This correlation suggests that amplitude strength in the input data influences the variability of diffusion model predictions, likely due to the multiplicative nature of noise corruption in these high-energy regions.
\end{itemize}

\subsection{Field data application}

\begin{figure*}[!b] 
\centering
\includegraphics[width=\textwidth]{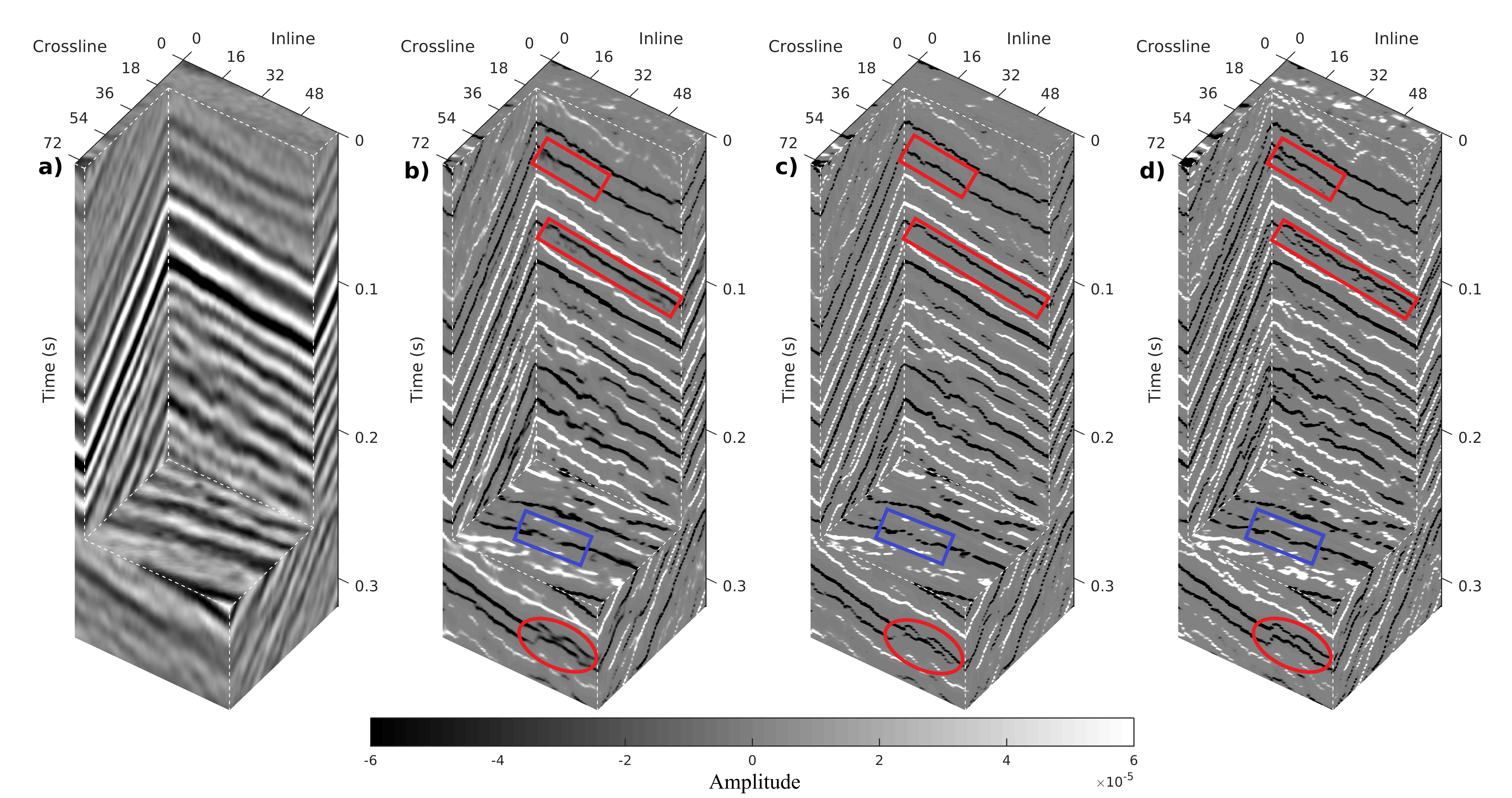}
\caption{Field data test. (a) Original low-resolution data. (b) SSI method result. (c) PISSD prediction. (d) Proposed method prediction.}
\label{fig11}
\end{figure*}

Having validated the method's performance on synthetic data under various noise conditions, we now evaluate its practical applicability on field seismic data. We apply the proposed method to a 3D post-stack seismic volume comprising 76 inline sections and 64 crossline sections. For computational efficiency, we extract 320 time samples from each trace at a 1 ms sampling interval for testing. Figure \ref{fig11}(a) shows the original low-resolution field data, where the amplitude distribution is uneven across depth, with weaker signals in the shallow and deep sections and less distinct event continuity compared to the middle section.

We estimate the seismic wavelet from the field data using geophysical software and apply three methods for resolution enhancement. Figure \ref{fig11}(b), (c), and (d) present results from SSI, PISSD, and the proposed method, respectively. All three methods improve the resolution of the original seismic data and balance the amplitudes of events at different depths. However, notable differences emerge in their detailed performance. In the red rectangular regions, the proposed method (Figure \ref{fig11}(d)) effectively identifies thin layers with weak energy, whereas SSI (Figure \ref{fig11}(b)) and PISSD (Figure \ref{fig11}(c)) fail to resolve these features. The blue rectangular regions highlight that the proposed method exhibits better event continuity. Additionally, in the red elliptical regions, the proposed method provides clearer and more accurate fault boundary information.

\begin{figure*}[!b] 
\centering
\includegraphics[width=\textwidth]{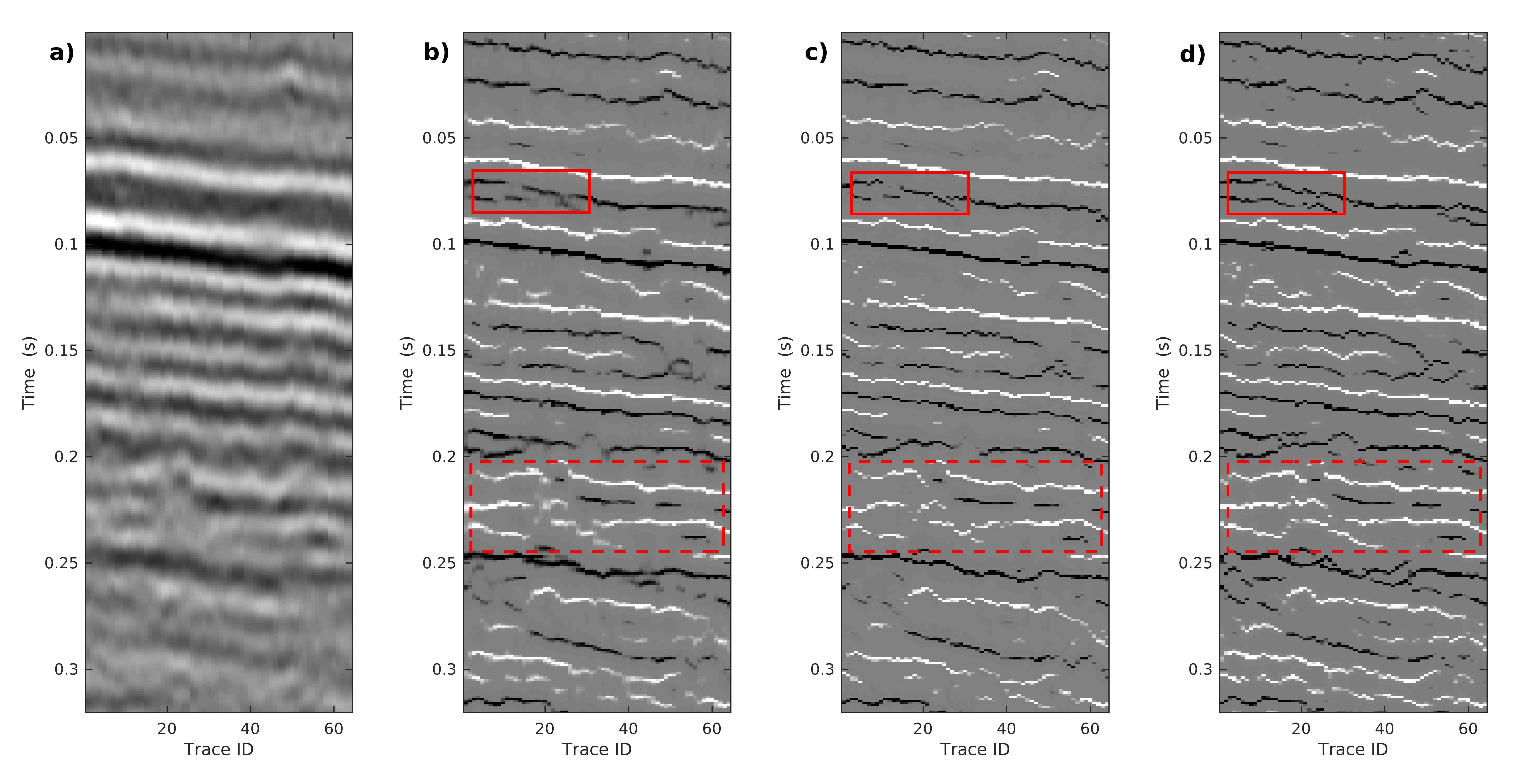}
\caption{ Zoomed-in view corresponding to inline \#5 in Fig.10. (a) Original low-resolution data. (b) SSI method result. (c) PISSD prediction. (d) Proposed method prediction.}
\label{fig12}

\vspace{0.2cm}

\centering
\includegraphics[width=\textwidth]{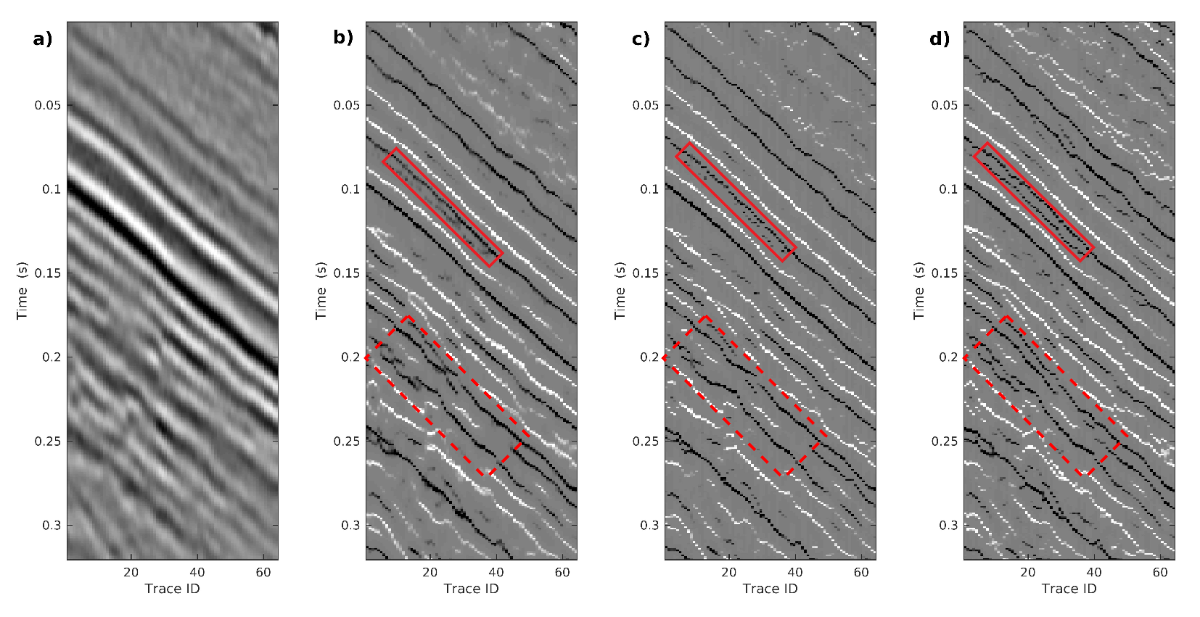}
\caption{ Zoomed-in view corresponding to crossline \#18 in Fig.10. (a) Original low-resolution data. (b) SSI method result. (c) PISSD prediction. (d) Proposed method prediction.}
\label{fig13}
\end{figure*}

For detailed examination, Figures \ref{fig12} and \ref{fig13} show the fifth inline profile and the 18th crossline profile, respectively. The red solid-line rectangles in both figures indicate thin-layer zones. While SSI and PISSD fail to accurately distinguish adjacent layers and exhibit poor continuity, the proposed method yields clear and continuous layers, demonstrating superior thin-layer recovery capability. The red dashed rectangles indicate deep and geologically complex areas where the original data exhibit weak amplitude and unclear reflector continuity. After processing, all three methods effectively balance the reflector amplitudes in these regions. However, the proposed method produces clearer structural trends and morphology than SSI and PISSD, demonstrating significant advantages in handling complex geological conditions.

\begin{figure*}[!t] 
\centering
\includegraphics[width=\textwidth]{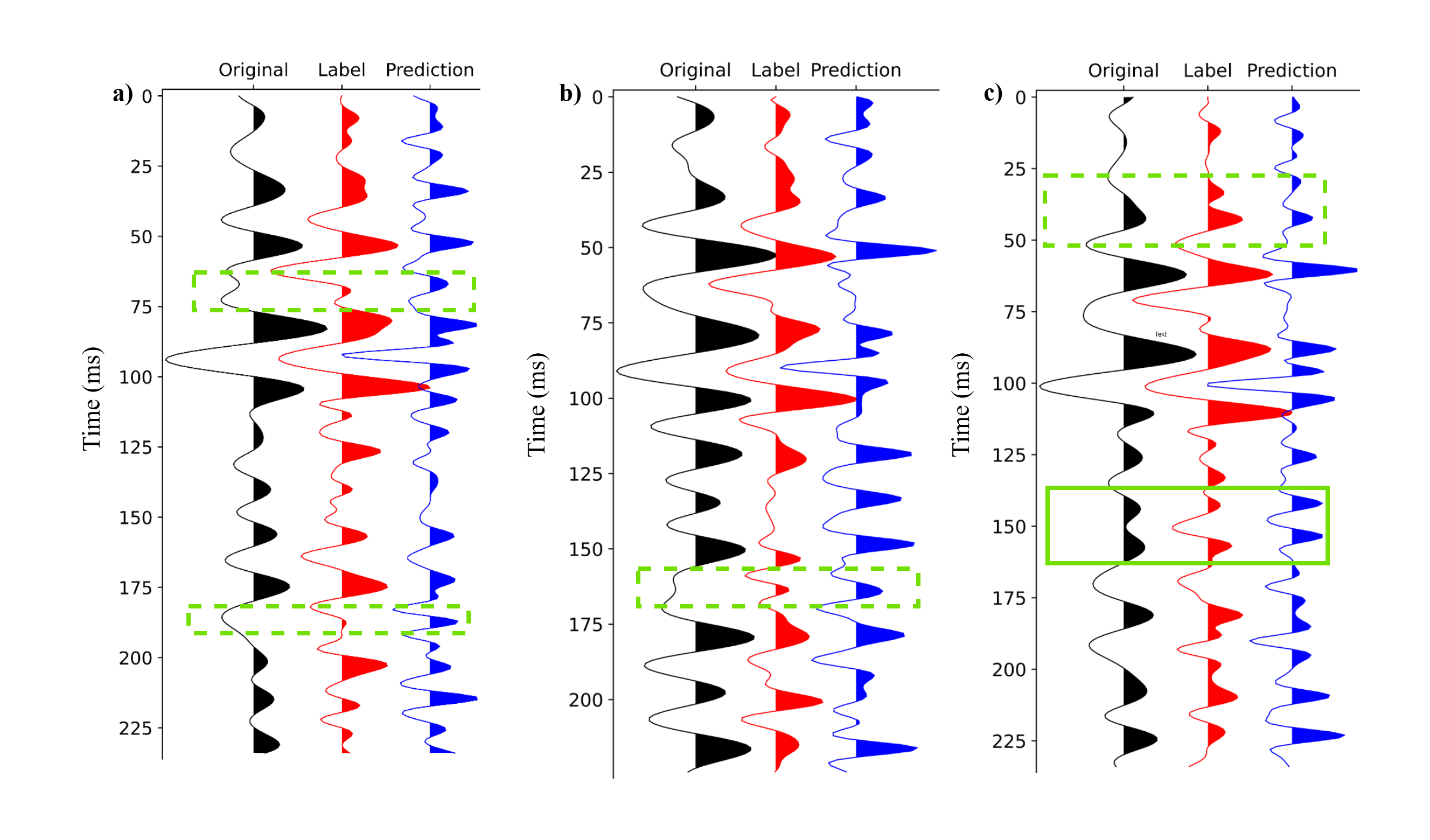}
\caption{Comparing the original low resolution seismic data (black line) and our predictions (blue line) with the well data (red line). (a)–(c) Wells at different locations.}
\label{fig14}
\end{figure*}

To verify the reliability of the enhanced results, we compare them against well log data at three well locations (Figure \ref{fig14}). In each subplot, the black line represents the original low-resolution seismic data, the blue line shows synthetic seismograms generated by convolving the high-resolution output from our method with a 0-150 Hz wavelet, and the red line displays synthetic seismograms from well log reflectivity convolved with the same wavelet. The comparison demonstrates strong agreement between the predicted results and well log synthetics, with notably better waveform consistency than the original seismic data (green solid rectangles). More importantly, the enhanced results reveal multiple thin layers (green dashed rectangles) that are absent in the original data but clearly present in the well log synthetics. This confirms that the proposed method reliably improves resolution and successfully recovers weak thin-layer features that were obscured in the original observations.

\begin{figure}[!t] 
\centering
\includegraphics[width=0.6\textwidth]{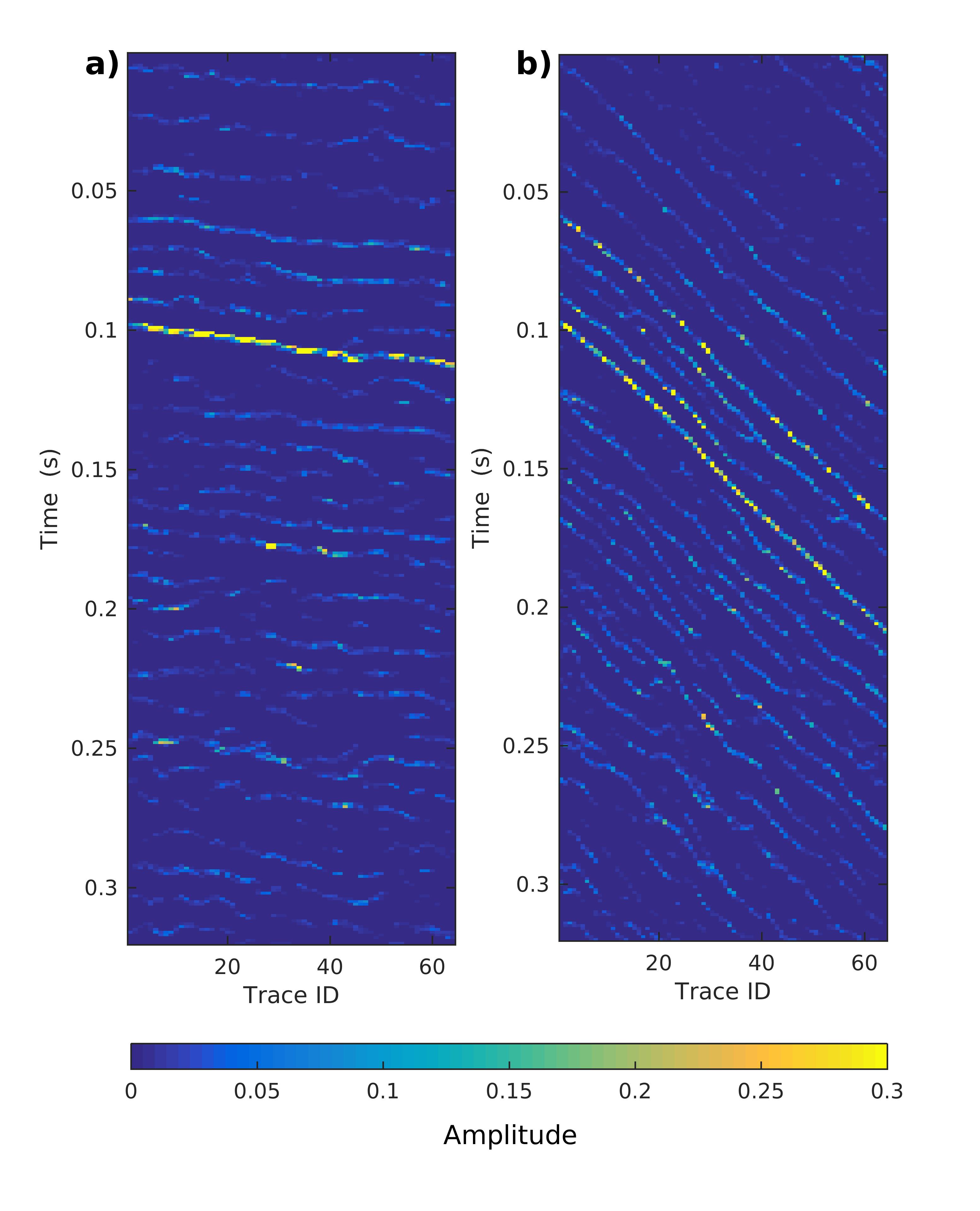}
\caption{Normalized standard deviation map of the 10 predictions using our method. (a) Inline \#5 profile. (b) Crossline \#18 profile.}
\label{fig15}
\end{figure}

Similarly, we also conducted an uncertainty analysis on the predictions made by the proposed method for the field data. Figures \ref{fig15}(a) and (b) show the normalized standard deviation distributions for inline \#5 and crossline \#18, respectively. We can see that regions with higher uncertainty exhibit spatial correlation with strong-amplitude zones in the original seismic data. This pattern indicates that amplitude variations in the original low-resolution data influence prediction reliability, with stronger amplitudes generally corresponding to elevated uncertainty. This uncertainty analysis offers valuable guidance for subsequent exploration decisions, such as identifying areas requiring additional data acquisition or prioritizing zones for detailed interpretation.

\section{Discussion}
The experimental results presented in Section \ref{examples} demonstrated that the proposed method achieves superior performance in seismic resolution enhancement compared to existing methods. To provide deeper insights into the framework, we examine four key aspects: the role of physical priors and progressive learning in enabling successful resolution enhancement, the benefits of physics-guided sampling for accelerating convergence, the contributions of multi-term loss functions in balancing different reconstruction objectives, and the limitations that suggest directions for future research. This discussion clarifies why the method works effectively and identifies opportunities for further improvements.

\begin{figure}[!t] 
\centering
\includegraphics[width=0.8\textwidth]{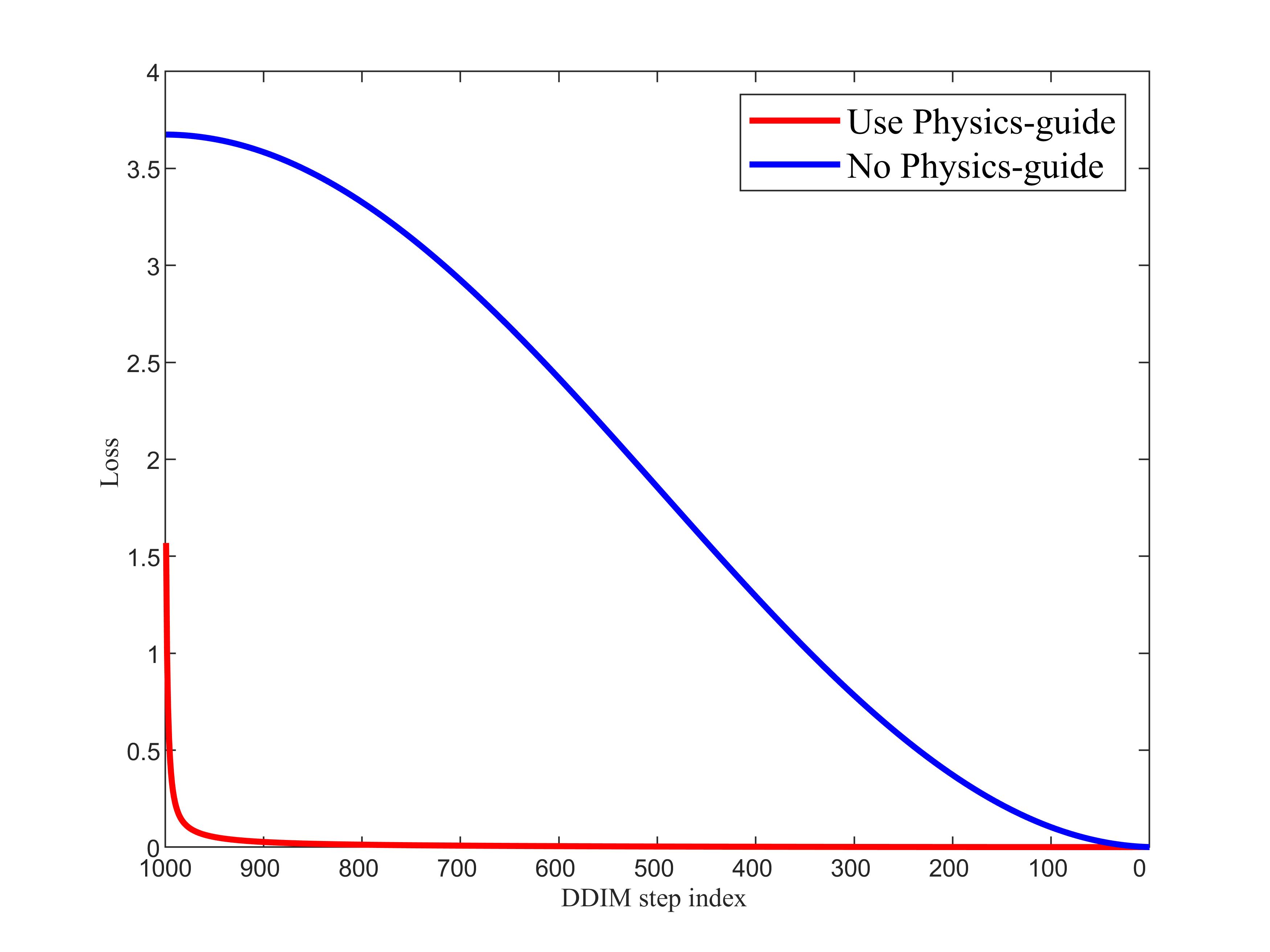}
\caption{Loss evolution during DDIM sampling: physics-guided (red) versus unguided (blue) cases.}
\label{fig16}
\end{figure}

\begin{figure}[!t] 
\centering
\includegraphics[width=\textwidth]{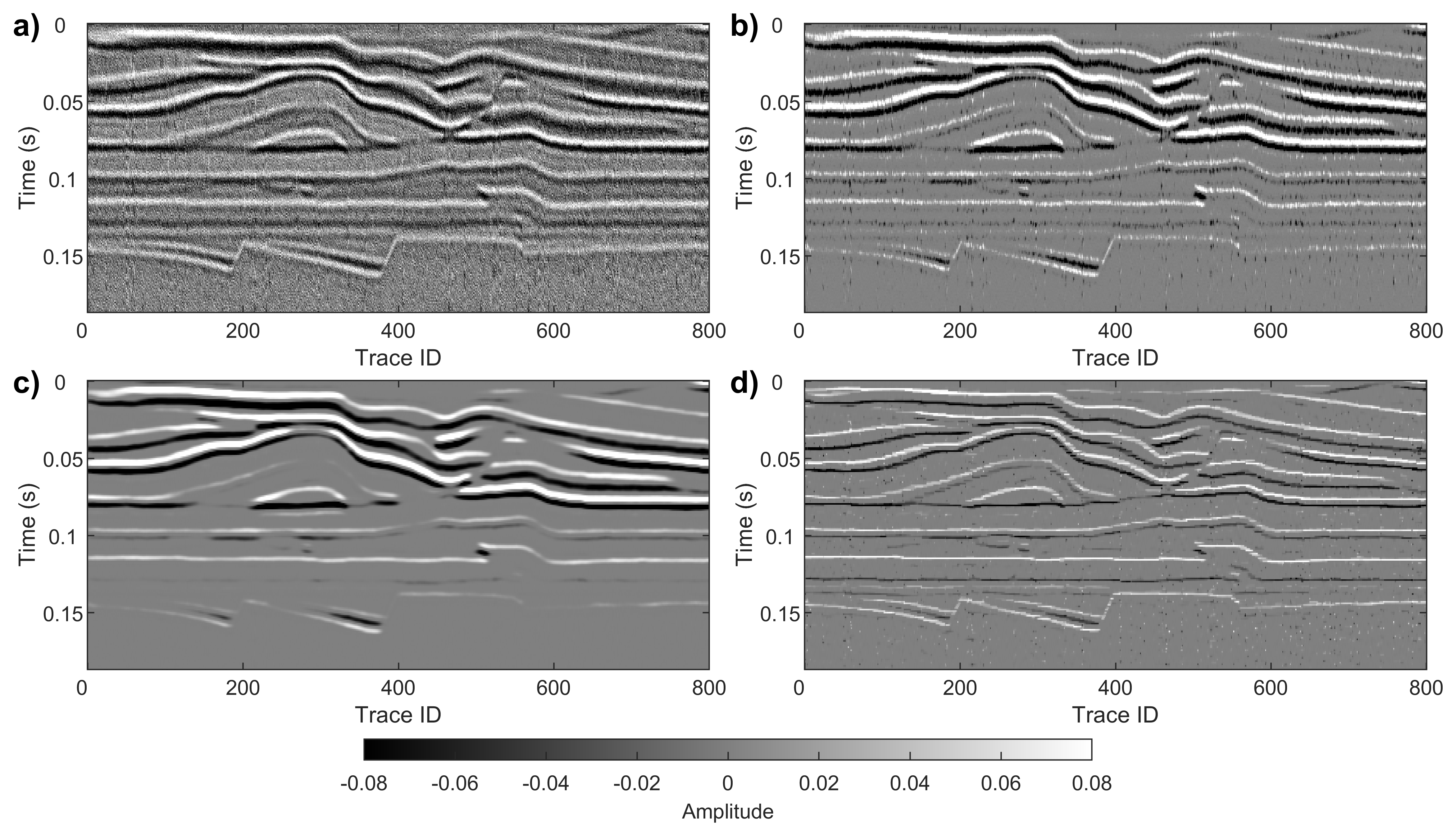}
\caption{Network performance under different loss functions. (a) Only use data loss (i.e., $\mathcal{L}_{\mathrm{mse}}$). (b) Data loss + sparsity loss ($\lambda_{\mathrm{mse}} \mathcal{L}_{\mathrm{mse}} + \lambda_{L_1} \mathcal{L}_{L_1}$). (c) Data loss + sparsity loss + total variation regularization (i.e., $\lambda_{\mathrm{mse}} \mathcal{L}_{\mathrm{mse}} + \lambda_{L_1} \mathcal{L}_{L_1} + \lambda_{\mathbf{t}_V} \mathcal{L}_{\mathbf{t}_V}$). (d) Data loss +  sparsity loss + total variation regularization + physical constraint loss (i.e., $\lambda_{\mathrm{mse}} \mathcal{L}_{\mathrm{mse}} + \lambda_{L_1} \mathcal{L}_{L_1} + \lambda_{\mathbf{t}_V} \mathcal{L}_{\mathbf{t}_V} + \lambda_{\mathrm{phy}} \mathcal{L}_{\mathrm{phy}}$).}
\label{fig17}
\end{figure}

\subsection{The role of physical priors and progressive learning}

The success of the proposed method stems from two fundamental design principles: the integration of physics-based prior knowledge and the progressive refinement strategy for distribution learning.

First, we include physics as prior knowledge for solution space constraint. Robinson's convolution model serves as a powerful prior that fundamentally constrains the solution space. The physical law $x = x_0 \ast w$ requires that any valid high-resolution output, when convolved with the wavelet and low-pass filtered, should reproduce the observed data. By embedding this constraint into both the training loss ($\mathcal{L}_{\text{phy}}$) and reverse sampling (gradient-based correction), the diffusion model learns to restrict predictions to a physically plausible manifold. This effectively reduces the vast space of possible outputs to a smaller subspace consistent with seismic physics. Unlike purely data-driven approaches that learn mappings solely from training examples, our method encodes governing physics directly into the optimization process. This ensures better generalization even with limited field data, as the model has learned underlying physical principles rather than just empirical patterns. The physical constraint acts as a regularizer that prevents physically implausible solutions, particularly important when dealing with noisy inputs.

Second, the progressive training strategy gradually shifts the target distribution from low to high resolution across multiple stages. In the initial stage ($k=0$), the diffusion model learns the distribution of the original observed seismic data. This provides a stable foundation, as the model first captures the inherent characteristics and structure of field seismic data at its native resolution. The model learns to represent the data manifold faithfully, including its geological structures, amplitude patterns, and spatial correlations. Subsequently, through progressive stages ($k=1, 2, \ldots, K$), we incrementally enhance the resolution of the training target. At each stage $k$, the enhanced output from stage $k-1$ becomes the new target, and its low-pass filtered version provides the conditioning input. This creates a sequence of increasingly challenging learning tasks, where the resolution gap between conditioning input and target output grows progressively larger. Therefore, we decomposes the difficult task of learning a large distribution shift into manageable smaller shifts. Each stage builds upon previous knowledge, allowing the model to refine its understanding incrementally. As a result, it allows the diffusion model to adapt its learned distribution representation step by step, moving smoothly along the manifold from low-resolution to high-resolution data distributions.

The combination of physics-based constraints and progressive distribution learning creates a synergistic effect. The physical prior ensures that at each progressive stage, the model moves toward physically consistent higher-resolution representations rather than arbitrary high-frequency patterns. Meanwhile, the progressive strategy provides a smooth learning trajectory that helps the model effectively incorporate physical constraints without getting trapped in poor local minima. Together, these principles enable the diffusion model to learn a high-quality mapping from low to high resolution that generalizes robustly to unseen field data.

\subsection{Acceleration through physics-guided sampling}

The physics-guided sampling process provides computational benefits by accelerating convergence to physically consistent solutions. Figure \ref{fig16} compares the evolution of physical loss $\mathcal{L}_{\text{phy}}$ during 1000-step sampling with and without physics guidance. Without physics guidance (blue curve), the physical loss decreases gradually throughout the entire sampling process, requiring all 1000 steps to reach low values. The model relies solely on learned priors to ensure physical consistency, which emerges slowly through progressive denoising. With physics-guided correction (red curve), the physical loss drops rapidly within the first 100 steps and converges to near-zero by approximately step 900. The gradient-based correction in Eq. (\ref{eq23}) actively steers each intermediate sample toward the physically consistent manifold, rather than relying on implicit enforcement through the full trajectory. This explicit correction immediately projects predictions onto the solution space satisfying Robinson's convolution model.

\subsection{The contributions of multi-term loss functions}

To understand the individual contributions of different loss components, we conduct an ablation study by progressively adding loss terms to the training objective. Figure \ref{fig17} shows results on synthetic noisy data (see Figure \ref{fig8}(a)) using four configurations: (a) data loss only ($\mathcal{L}_{\text{mse}}$), (b) data loss with sparsity constraint ($\mathcal{L}_{\text{mse}} + \mathcal{L}_{L_1}$), (c) adding total variation (TV) regularization ($\mathcal{L}_{\text{mse}} + \mathcal{L}_{L_1} + \mathcal{L}_{\text{tv}}$), and (d) the complete composite loss including physical constraints ($\mathcal{L}_{\text{mse}} + \mathcal{L}_{L_1} + \mathcal{L}_{\text{tv}} + \mathcal{L}_{\text{phy}}$).

Using only MSE loss (Figure \ref{fig17}(a)), the network produces noisy results with poor signal-to-noise ratio. While some geological structures are visible, substantial artifacts persist throughout the section. The lack of regularization allows the model to overfit to noise during training, resulting in the artifacts in the predictions. Adding $L_1$ sparsity loss (Figure \ref{fig17}(b)) suppresses noise to some extent and slightly improves the resolution of reflection events. The sparsity constraint promotes solutions with sparse reflectivity, reducing background noise. However, the improvement in resolution remains limited, and lateral continuity of events is still poor with visible discontinuities along reflectors. Incorporating TV regularization (Figure \ref{fig17}(c)) further improves lateral continuity of seismic events. The TV term penalizes abrupt spatial variations, encouraging smooth transitions along the horizontal direction and enhancing event coherence. However, resolution enhancement remains modest compared to expectations. More critically, some reflection events are significantly suppressed, as the TV regularization tends to oversmooth the data in its attempt to enforce spatial continuity. The complete composite loss with physical constraints (Figure \ref{fig17}(d)) yields the best results. The complete composite loss with physical constraints (Figure \ref{fig17}(d)) yields the best results. The addition of $\mathcal{L}_{\text{phy}}$ enforces consistency with Robinson's convolution model, ensuring that predictions satisfy fundamental seismic physics. This final constraint eliminates remaining artifacts and produces the sharpest, most continuous events with optimal signal-to-noise ratio.

While the composite loss function improves physical plausibility and stability, it introduces the challenge of tuning four weighting coefficients $\lambda_{\text{mse}}$, $\lambda_{\text{phy}}$, $\lambda_{L_1}$, and $\lambda_{\text{tv}}$. The high-dimensional parameter space complicates hyperparameter optimization. To address this, we adopt two strategies. First, we employ adaptive weighting during training, dynamically adjusting the contribution of each loss term to balance competing objectives as training progresses. Second, we use a phased tuning approach: initially setting higher $\lambda_{\text{mse}}$ to ensure stable convergence, then gradually increasing $\lambda_{\text{phy}}$ and the regularization weights $\lambda_{L_1}$ and $\lambda_{\text{tv}}$ to refine physical consistency and spatial quality. This staged strategy reduces the effective parameter search space and improves training stability.

\subsection{Limitations and future directions}

While the proposed method demonstrates superior performance, it has inherent limitations that warrant discussion and suggest directions for future research. The physics-guided framework relies on the availability of a reasonable seismic wavelet. Robinson's convolution model $x = x_0 \ast w$ explicitly requires the wavelet $w$ as prior knowledge to constrain the solution space and guide the sampling process. In the field data application, we estimated the wavelet using commercial geophysical software. While wavelet estimation is a standard procedure in seismic processing and can be performed for any dataset, the quality of estimation varies depending on data characteristics, signal-to-noise ratio, and the presence of well log constraints. Inaccurate wavelet estimates can degrade the effectiveness of physical guidance, potentially leading to biased resolution enhancement or artifacts.

This limitation raises an important question: can we relax the requirement for pre-estimated wavelets by incorporating wavelet estimation into the optimization framework itself? Two potential strategies merit investigation. First, joint optimization could simultaneously enhance resolution and refine the wavelet during training or inference. The diffusion model would treat both the high-resolution seismic data and the wavelet as unknowns to be recovered, potentially using additional constraints such as wavelet smoothness or causality. Second, an alternating optimization scheme could iteratively update resolution and wavelet: (1) fix the current wavelet estimate and optimize for high-resolution data, (2) fix the enhanced data and update the wavelet to better satisfy the convolution relationship, (3) return to step 1 with the refined wavelet for further resolution enhancement. This alternating approach could progressively improve both the data resolution and wavelet accuracy through multiple cycles.

Such joint or alternating optimization strategies would reduce dependence on external wavelet estimation tools and potentially improve robustness when initial wavelet estimates are poor. However, they also introduce additional complexity, including convergence challenges and the risk of non-unique solutions when both unknowns are optimized simultaneously. Careful regularization and initialization would be essential. Exploring these extensions represents a promising direction for future work that could enhance the method's applicability to datasets where reliable wavelet estimation is challenging.

\section{Conclusions}\label{sec:conclusions}
We proposed a novel physics-guided self-supervised diffusion model for seismic resolution enhancement. The method integrated the generative capability of diffusion models, the flexibility of self-supervised learning, and the physical principles of seismic convolution models, establishing a processing framework that achieves both high fidelity and strong generalization. By adopting a progressive iterative training strategy combining low-pass filtering and high-resolution prediction, the method eliminated the dependence on paired labeled data while maintaining physical consistency through Robinson's convolution model. The incorporation of a composite loss function during training further ensured that the enhanced results satisfied data fidelity, physical constraints, sparsity priors, and spatial regularity. Additionally, the integration of physics-guided correction in the reverse sampling process guaranteed that the generated high-resolution seismic data respected the convolution relationship at each denoising step.

Experimental results on both synthetic and field data demonstrated the superior performance of the proposed method compared to SSI and PISSD. The method effectively improved the resolution and lateral continuity of seismic events, enabling more accurate characterization of thin layers and subtle geological structures. Under challenging conditions such as strong random noise, realistic noise, and amplitude imbalance in complex geological settings, the method maintained stable performance and robust noise suppression. The uncertainty quantification capability provided by the diffusion framework offered valuable guidance for risk assessment in exploration decisions. As a result, this study provided a physics-data integrated technical approach for high-resolution reconstruction of seismic signals. The results demonstrated significant potential for supporting high-precision reservoir interpretation and inversion tasks in practical exploration applications.

\section*{Acknowledgments}
This work is supported by the National Natural Science Foundation of China (Nos. 42404140, 42130808). The author Shijun Cheng thanks King Abdullah University of Science and Technology for supporting this research. The authors thank Qianru Xu for the valuable discussions and the BGP Research and Development Center for graciously supplying the field data.

\bibliographystyle{unsrtnat}
\bibliography{references}

\end{document}